\newcommand*{\ch}{} 
\newcommand*{\chf}{} 

\newcommand*{\chs}{} 
\newcommand*{\chn}{} 
\newcommand*{\che}{} 

\newcommand{\chni}{}
\newcommand{\cht}{}
\newcommand{\chel}{}

\documentclass[twocolumn]{aastex62}

\usepackage{amsmath,amstext}
\usepackage{gensymb}

\def\mJybeam{mJy\,beam$^{-1}$}

\def\as{\farcs}

\def\DCOp{DCO$^{+}$}
\def\C18O{C$^{18}$O}
\def\3CO{$^{13}$CO}
\def\2CO{$^{12}$CO}

\newcommand\aastex{AAS\TeX}
\newcommand\latex{La\TeX}

\graphicspath{{./}{figures/}}

\received{March 29, 2019}
\submitjournal{ApJ}

\shorttitle{{\che The} physical conditions of VLA1623A circumbinary disk}
\shortauthors{Hsieh et al.}

\begin{document}

\title{Determining the physical conditions of extremely young Class 0 circumbinary disk around VLA1623A}

\correspondingauthor{Cheng-Han Hsieh}
\email{cheng-han.hsieh@yale.edu}

\author[0000-0003-2803-6358]{Cheng-Han Hsieh}
\affil{Department of Physics,
National Tsing Hua University,101 Section 2 Kuang Fu Road,
30013 Hsinchu, Taiwan}
\affiliation{Department of Astronomy, Yale University, New Haven, CT 06511, USA}

\author[0000-0001-5522-486X]{Shih-Ping Lai}
\affil{Department of Physics,
National Tsing Hua University,101 Section 2 Kuang Fu Road,
30013 Hsinchu, Taiwan}
\affiliation{Institute for Astronomy,
National Tsing Hua University,101 Section 2 Kuang Fu Road,
30013 Hsinchu, Taiwan}
\affiliation{Academia Sinica Institute of Astronomy and Astrophysics, PO Box 23-141, 10617 Taipei, Taiwan}
\author{Pou-Ieng Cheong}
\affiliation{Institute for Astronomy,
National Tsing Hua University,101 Section 2 Kuang Fu Road,
30013 Hsinchu, Taiwan}

\author[0000-0003-2158-8141]{Chia-Lin Ko}
\affiliation{Institute for Astronomy,
National Tsing Hua University,101 Section 2 Kuang Fu Road,
30013 Hsinchu, Taiwan}

\affiliation{Department of Physics, National Taiwan Normal University,
              No.162, Sec.1, Heping East Road, Taipei 10610 , Taiwan}

\author{Zhi-Yun Li}
\affiliation{Department of Astronomy,University of Virginia,
            530 McCormick Road Charlottesville VA, USA}

\author{Nadia M. Murillo}
\affiliation{Leiden Observatory, Leiden University, PO Box 9513, 2300 RA Leiden, The Netherlands}



\begin{abstract}


We {\cht present detailed analysis of high resolution C$^{18}$O (2-1), SO ($8_{8}$-$7_{7}$), CO (3-2) and DCO$^{+}$ (3-2) data obtained by the Atacama Large Millimeter/sub-millimeter Array (ALMA) towards a Class 0 Keplerian circumbinary disk around VLA1623A, which represents one of the most complete analysis towards a Class 0 source. From the dendrogram analysis, we identified several accretion flows feeding the circumbinary disk in a highly anisotropic manner. Stream-like SO emission around the circumbinary disk reveals the complicated shocks  caused by the interactions between the disk, accretion flows and outflows. A wall-like structure is discovered south of VLA1623B. The discovery of two outflow cavity walls at the same position traveling at different velocities suggests the two outflows from both VLA1623A and VLA1623B overlays on top of each other in the plane of sky. Our detailed flat and flared disk modeling shows that Cycle 2 C$^{18}$O J = 2-1 data is inconsistent with the combined binary mass of $0.2\,M_\odot$ as suggested by early Cycle 0 studies. The combined binary mass for VLA1623A should be modified to $0.3\sim0.5\,M_\odot$.}




\end{abstract}

\keywords{star formation --
                individual objects: VLA1623 --
                accretion disks -- methods: observational -- stars: low-mass --techniques: interferometric
               }


\section{Introduction} 
\label{sec:intro}

{\che Around 50 percent of solar mass stars form in multiple systems \citep{2010ApJS..190....1R}. Multiplicity fraction increases for higher mass stars \citep{2011IAUS..272..474S}. These multiple systems are formed in the early stage of star formation via three processes: turbulent fragmentation, thermal fragmentation of rotating cores, and disk fragmentation. Turbulent and thermal fragmentation occurs {\cht at} relatively large scales, forming wide binaries with separation of order 1000 AU or larger.  \citep{2002ApJ...576..870P,2010ApJ...725.1485O,1992ApJ...388..392I,1993MNRAS.264..798B,2014ApJ...794...44B,2015Natur.518..213P}. As for the disk fragmentation, it is believed to be one of the primary processes for forming close  ($\sim$ 100 AU) binaries \citep{2012ApJ...754...52T,2013ApJ...771...48T}}. 



Previous observations have found many circumbinary disks \citep{2012ApJ...754...52T,2013ApJ...771...48T,2014Natur.514..600D,2016A&ARv..24....5D,2014ApJ...796...70C,2014ApJ...793...10T,2016ApJ...820...19T}. Near-infrared surveys of Class I sources have found that around 15 out of 88 targets have binary separations between 50 to 200 AU \citep{2008AJ....135.2526C,2007A&A...476..229D}. {\cht Very Large Array survey of 94 known protostars in Perseus molecular clouds have found that Class 0 have significant higher multiplicity fraction (MF$=0.57 \pm 0.09$) as compared to Class I sources (MF$=0.23 \pm 0.08$)\citep{2016ApJ...818...73T}. Submillimeter Array (SMA) studies of 33 Class 0 protostars in the nearby molecular clouds also found the multiplicity fraction (MF$=0.64 \pm 0.08$) to be two times larger than Class I sources \citep{2013ApJ...768..110C}.} Since disk fragmentation requires massive gravitationally unstable disks, close binary and multiplicity systems are expected to form in the early phase of star formation. Thus understanding the gas dynamics in extremely young Class 0 protobinary disk is crucial for testing binary formation theory. 

Disk formation in the Class 0 phase {\chel attracted} a lot of attention over the last 10 years and it remains an important unsolved problem for star formation. Numerical models simulating the collapse of a magnetized envelope with the assumption of ideal magnetohydrodynamics (MHD) show that disk formation is hindered by magnetic braking effects \citep{2008ApJ...681.1356M}. One solution proposed is that magnetic braking efficiency can be reduced if the rotation axis is misaligned with the magnetic field direction \citep{2009A&A...506L..29H}. Moreover, non-ideal MHD effects on disk formation {\che have} also been explored. Recently, 3D non-ideal MHD {\che simulations have} been carried out and a disk around 5\,AU is {\che formed} at the end of the first core phase \citep{2015ApJ...801..117T}. {\cht In the recent analytical study carried out by \cite{2016ApJ...830L...8H}, a relationship between disk radius and magnetic fields in the inner part of the core is found. The weak dependence of various relevant quantities suggests that Class 0 disks have a typical disk size $\sim 18$\,AU \citep{2016ApJ...830L...8H}.}


{\cht An alternative solution to magnetic braking effects is turbulence. \citet{2012ApJ...747...21S,2013MNRAS.429.3371S} suggests that turbulence reconnection in disks is associated with the loss of magnetic flux which reduces the magnetic braking efficiency. In contrast, \citet{2015MNRAS.446.2776S} have shown even with mild sub-sonic turbulence motion together with disorder magnetic field is enough for the formation of Class 0 Keplerian disks without the need for the loss of magnetic flux. In their simulations they found that both the accretion of mass and angular momentum is highly anisotropic.}




{\cht Numerical simulations tend to have difficulties in explaining how the observed large Class 0 disks grow in size and possibly fragment.} A well-studied Class 0 candidate, HH212 is found to have a Keplerian disk within the 44 AU centrifugal barrier  \citep{2017ApJ...843...27L}. The Keplerian disk of Class 0 protostar L1527 has a disk radius of 74 AU and the scale height at 100 AU is 48 AU  \citep{2017ApJ...849...56A,2013ApJ...771...48T}. As for VLA1623A, the Keplerian rotating disk size is fitted to 150 AU using C$^{18}$O as a tracer \citep{2013A&A...560A.103M}. These observational results show Class 0 disks with varying disk {\che sizes and structures}. {\cht The discovery of large ($>$ 70AU) Class 0 rotationally supported disks, L1527 and VLA1623A, provide possible candidates to study disk fragmentation around Class 0 sources. Not much is known about how large rotationally supported {\che disks are} formed, and what {\che factors influence} their formation.} 

\begin{figure}[tbh!]
\centering
\includegraphics[width=\hsize]{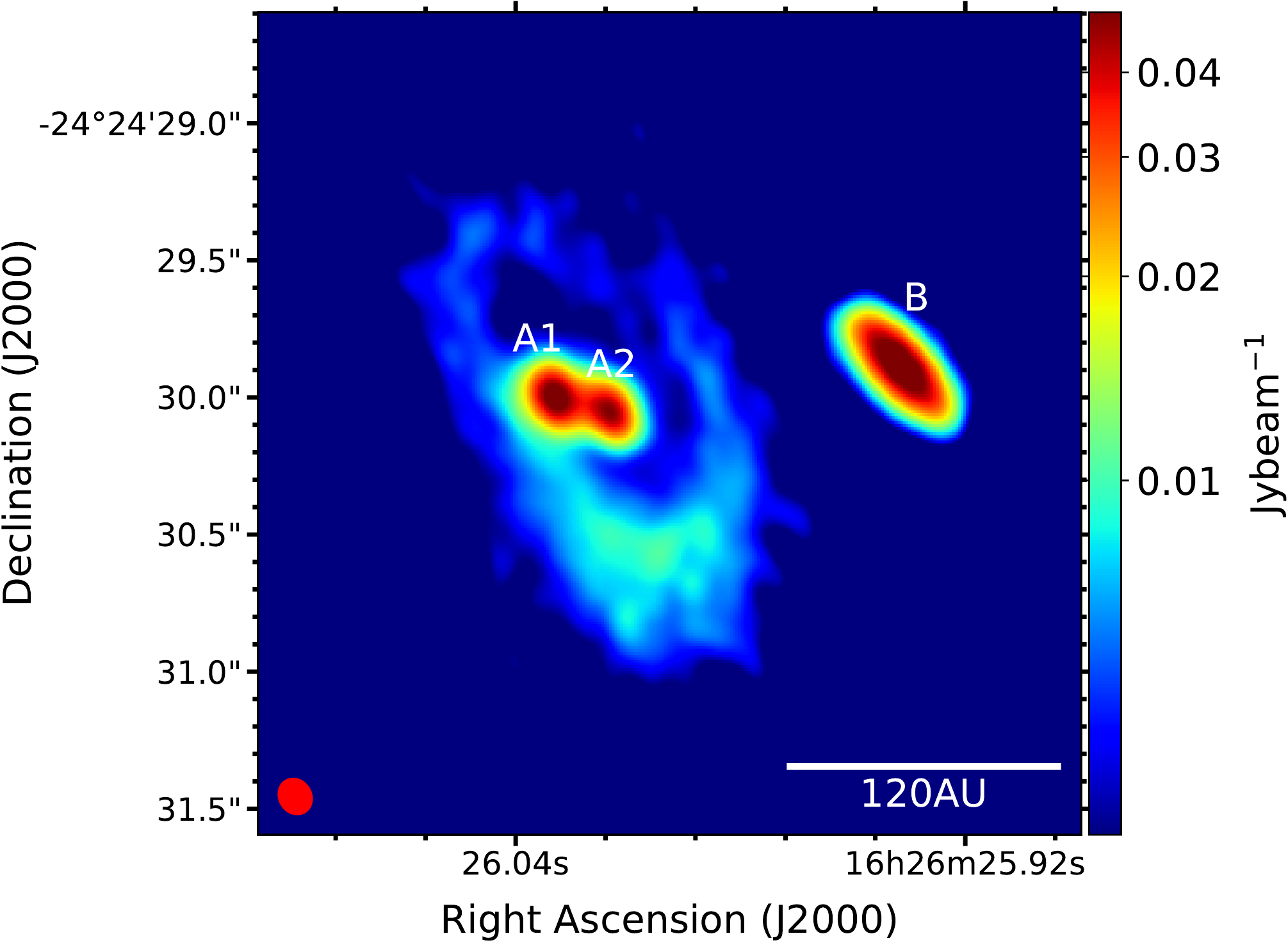}
\caption{VLA1623 0.88\,mm continuum image. {\che The Class 0 source, VLA1623A is resolved into 2 sources, VLA1623A1 and VLA1623A2. The red ellipse marks the beam size.}
}
\label{fig1}
\end{figure} 
 
\begin{figure*}[tbh]
\centering
\makebox[\textwidth]{\includegraphics[width=\textwidth]{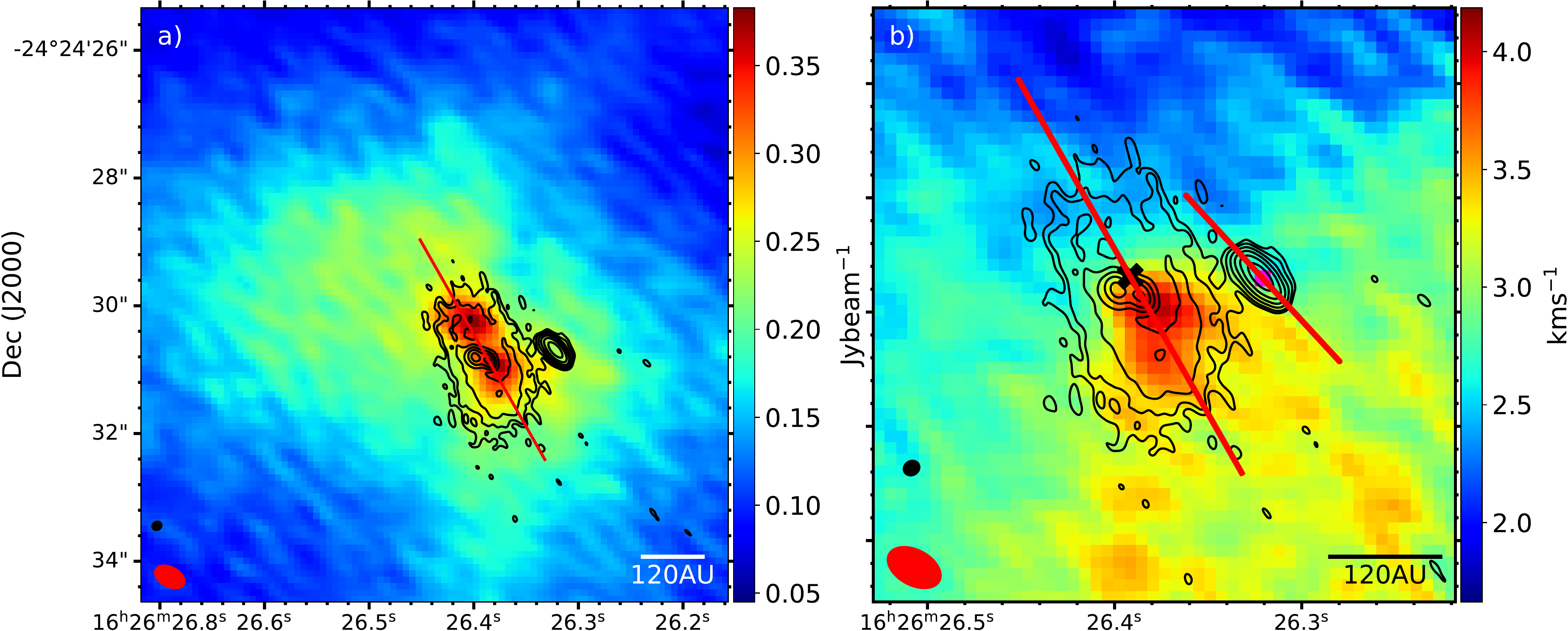}}
\caption{(a).VLA1623 0.88\,mm continuum (contours) overlaid on the \textbf{\chni Briggs -1.0 weighted} \C18O integrated intensity map (color). The color represents the C$^{18}$O J = 2-1 integrated intensity map (moment 0) using combined data from {\che ALMA} 12m array and 7m array. The black contours are 0.88 mm continuum data in steps of 3\,$\sigma$, 5\,$\sigma$, 10\,$\sigma$, 20\,$\sigma$, 40\,$\sigma$, 80\,$\sigma$, where $\sigma = 5\times 10^{-4}$ Jy beam$^{-1}$. The red line indicates the location of the position velocity cuts used in this study (PA = 209.82$^{\circ}$, centered at VLA1623A {\cht ($\alpha$(J2000) = 16\textsuperscript{h}26\textsuperscript{m}26\fs390, $\delta$(J2000) = --24\degree24\arcmin30\as688);} PA=222.80$^{\circ}$, centered at VLA1623B {\cht ($\alpha$(J2000) = 16\textsuperscript{h}26\textsuperscript{m}26\fs305, $\delta$(J2000) = --24\degree24\arcmin30\as705)}).  (b). VLA1623 0.88\,mm continuum (contours) overlaid on the Briggs -1.0 weighted \C18O intensity weighted velocity map (color). VLA1623A and B's position are marked with a cross and a square, respectively.
}
\label{fig2}
\end{figure*}

VLA1623-2417 (VLA1623 hereafter) is a triple non-coeval protostellar system, located at {\chn 120$\pm$4.5}\,pc away in cloud $\rho$ Ophiuchus A \citep{1993ApJ...406..122A,2008ApJ...675L..29L,2017ApJ...834..141O}\footnote{In this paper, the distance of 120\,pc \citep{2013ApJ...764L..15M} is used in the modelling instead of 137.3\,pc \citep{2017ApJ...834..141O}. The physical scale would increase by a factor of 1.14 if the 137.3\,pc distance is adopted.}. The system {\che consists} of three components: VLA1623A (Class 0 source), VLA1623B {\che (younger than VLA1623A, possibly a transition between starless core and Class 0) \citep{2015A&A...581A..91S}}, and VLA1623W (Class 1) \citep{2013ApJ...764L..15M}. 
{\chn Recent} high angular resolution ($\sim$0\as16) 0.88 mm continuum data reveals {\che that} VLA1623A is in fact a binary  denoted as VLA1623A1 and VLA1623A2 in \autoref{fig1}, and the Keplerian disk previously discovered is a circumbinary disk \citep{2018ApJ...861...91H}. The plane of sky separation between the {\che VLA1623A1 and VLA1623A2}\footnote{\che The nomenclature used in Harris+2018 is VLA1623Aa and VLA1623Ab. However, the lower case letters after a source name are reserved for planets. IAU stipulated convention for stars should be VLA1623A1 and VLA1623A2.} is around 0\as2, which corresponds {\chs to a physical scale of} 24 AU. Unfortunately, {\che our} Cycle 2 C$^{18}$O (J = 2-1) data with beam size 0\as5 {\che cannot} resolve the two sources (C$^{18}$O Moment maps shown in \autoref{fig2}). 


\begin{table*}[t]
\setlength{\tabcolsep}{6pt} 
\caption{Summary of the observational data used in the analysis}             
\label{table:1}      
\centering                          
\begin{tabular}{c c c c c c c c}        
\hline\hline                 
 &  &   &   & {\chni Maximum} &Channel   &  &  \\
{\chni Line} & {\chni Transition} &{\chni $\nu$} &{\chni Beam Size} &{\chni Recoverable} &{\chni Widths} &Rms noise&PI \\    
& &(GHz) &(\as) &{\chni Scale (\as)} &(kms$^{-1}$) &(mJy\,beam$^{-1}$)& \\ 
\hline                        
{\chni CO} & {\chni 3-2} & {\chni 345.79599} &{\chni $0\as99 \times 0\as59$ }&{\chni $ 4\as33$}& {\chni 0.0529 }& {\chni 11}& {\chni Cheng-Han Hsieh}\\

   C$^{18}$O & 2-1 & 219.56036& $0\as52 \times {\ch 0\as31}$ &{\chni $ 23\as26$}& 0.0832& {\ch 9}& \\      
    &  &    & $1\as10 \times 0\as89$ &{\chni $ 23\as26$}& 0.0208 & {\ch 18} &Shih-Ping Lai\\      
   &  &    & $0\as51 \times 0\as31$ &{\chni $ 23\as26$}& 0.0208 & {\ch 14} & \\

   DCO+ & {\ch 3-2} & {\ch 216.11402 }  & $0\as51 \times 0\as31$ &{\chni $ 23\as26$}& 0.085& 5 &Shih-Ping Lai \\ 
   
   SO &$\nu=0$, & 344.31061  &$1\as11 \times 0\as76$ &{\chni $ 3\as9$}& 0.212 & 20 &{\che Victor}  \\ 
   &$J = 8_{8}-7_{7}$&&&&&&{\chni Magalh$\tilde{a}$es}\\
   Continuum & ... & 336.50000   & $0\as13 \times 0\as12$ &{\chni $ 0\as88$}& 53309.11 & 0.5 &{\che Leslie Looney} \\ 
   {\chf Total Power Array}&{\chf ...}&{\chf 219.56036}&{\chf $29\as67 \times 29\as67$}&{\chni 427\as86}&{\chf 0.0208}&{\chf 500}&{\chf Shih-Ping Lai}\\
  
\hline                                   
\end{tabular}
\end{table*}
With the increase of antennas and the UV-coverage, Cycle 2 data picks up significantly more C$^{18}$O emission than Cycle 0 observations shown in \autoref{fig3}. VLA1623A, being one of the youngest protobinary disks ever discovered is thus the perfect candidate to study disk fragmentation around Class 0 sources. 

In what follows, in Section \ref{sec:Observation} we discuss the data used in this analysis. In Section \ref{sec:3.1} we compare Cycle 0 and Cycle 2 C$^{18}$O data and highlight major emission components that would be further explored. In Section \ref{sec:analysis} we present our main results of {\cht large scale emission} fitting, dendrogram analysis (Cheong et al. {\chni 2019; submitted}), flat and flared disk modeling, and accretion shocks analysis. In Section \ref{sec:discussion}, we summarized the physical conditions of the VLA1623A circumbinary disk and its surrounding environment. In Section \ref{sec:conclusion}, we give our conclusions. 



\section{{\che Observations}}
\label{sec:Observation}

\subsection{{\chni CO}}

{\chni We observed the CO ($J\,=\,3-2$) emission with Atacama Large Millimeter/submillimeter Array (ALMA) in Cycle 6 with pointing coordinates ($\alpha$(J2000) = 16\textsuperscript{h}26\textsuperscript{m}26\fs390, $\delta$(J2000) = --24\degree24\arcmin30\as688). The observation was taken on 2019 March 15 using 47 antenna of 12\,m array in C34-1 compact configuration {\chni (project ID: 2018.1.00388S, PI: Chenghan Hsieh)}. The total on-source integration time is 10 minutes sampling baseline ranges between 15.1 $\sim$ 360.6\,m. ALMA pipeline and Common Astronomy Software Applications (CASA) version 5.4.0-70 is used to calibrate the visibility data with J1924+2914 as the calibrator for bandpass and flux calibration and J1625+2527 for phase calibration. We used CASA CLEAN task with briggs 0.5 weighting and Hogbom as deconvolver for imaging. The resulting image has a beam size of 0\as99 $\times$ 0\as59 (P.A. = -84.8\degree) with rms noise level at 11 \mJybeam and velocity resolution at 0.0529\, kms$^{-1}$.}

\subsection{C$^{18}$O}
C$^{18}$O ($J = 2-1$) were observed with ALMA in Cycle 2 with 12\, m array configuration C34-5 and C34-1 {\chni (project ID: 2013.1.01004.S, PI: Shih-Ping Lai)}. We also include the 7m Atacama Compact Array (ACA) (hereafter 7m array) and Total Power Array of ACA in our analysis. {\che The} 12m array data and 7m array data are combined via the CASA task CLEAN with a weighting parameter of {\che Briggs -1.5, Briggs -1.0, and a natural weighting.} {\che These maps are used for identifying accretion flows, comparing with ALMA C$^{18}$O Cycle 0 data, and analysis of disk motion respectively.} 
 


For the Briggs {\chs -1.5} weighting, the UV taper range is $0\as25$. The {\che resulting} C$^{18}$O (J = 2--1) channel map {\che has a} resolution of $1\as1\times 0\as89$ (P.A. = $36.8^{\circ}$) with rms noise level at 18 \mJybeam. The velocity resolution is 0.0208 kms$^{-1}$ with rest frequency at 219.56036 GHz (corresponding to {\che a} system velocity of 4.0 kms$^{-1}$ away in the line of sight). The low resolution data is used for {\cht the} dendrogram analysis and for identifying large structures around the circumbinary disk {\cht VLA1623A}.


   
For the natural weighted data, Hogbom algorithm is used for {\chs the} deconvolution {\chs process resulting the resolution of} $0\as52 \times 0\as31$ (P.A. = $87.89^{\circ}$) with rms noise level at {\ch 9} mJy beam$^{-1}$. The velocity resolution {\chs of} 0.0832 kms$^{-1}$ {\chs is used} to achieve a higher signal to noise ratio {\chs for disk modeling.} 
 
{\chs For comparison purposes, Briggs -1.0 weighting with UV taper range $0\as25$ is applied to the Cycle 2 C$^{18}$O data to achieve resolution $0\as51 \times 0\as31$ (P.A. = $61.83^{\circ}$) with rms noise level at 14 mJy beam$^{-1}$. The velocity resolution is 0.0208 kms$^{-1}$. This {\che data set} is {\chni used} to compare with Cycle 0 observation in \autoref{fig3}.} 
   
The C$^{18}$O mean velocity map (\autoref{fig2} b.) shows a rotating disk around VLA1623A. In this study, all the position-velocity (PV) diagrams will be {\chni aligned} along the red line in \autoref{fig2}. The combined {\chs natural weighted} C$^{18}$O data for {\chs 12m array configuration} C34-5 (baseline range $18.9 \sim 1090$ m), C34-1 (baseline range $14.2 \sim 347.3$ m), and 7m array (baseline range $7.1 \sim 48.9$ m) data will be used in the analysis. 

As for the Total Power Array data, the angular resolution is $29\as67 \times 29\as67$ (P.A. = $0.0^{\circ}$) with rms noise level at 0.5 Jy beam$^{-1}$. The script provided by ALMA Regional Center (ARC) is used to calibrate and reduce the data. The Total Power Array data would be used {\cht to identify} the {\cht large scale emission in the foreground as well as the background} of the VLA1623A circumbinary disk. 

\subsection{Continuum data}
We use the high resolution continuum data from the ALMA Archive (project ID: 2015.1.00084.S, PI: Leslie Looney) {\ch see \autoref{table:1}}. The observation is done in Band 7 with an integration time of 4798 seconds. The continuum data is prepared by using CASA task CLEAN with uniform weighting. The synthesized beam is {\ch $0\as13 \times 0\as12$} (P.A. = $31.00^{\circ}$) with rms noise level at 0.5 mJy beam$^{-1}$. 

\subsection{\DCOp $ J= 3-2$ data}
 The DCO+ data used in this paper {\cht is} from ALMA Cycle 2 observation in Band 6 of the 12m array combined with the observation from the 7m array {\chni (project ID: 2013.1.01004.S, PI: Shih-Ping Lai)}. The data is prepared by the CASA task CLEAN with a weighting parameter of robust -0.5, and the UV taper range is set to 1\as0. The rms noise level of DCO+ data is $5$ mJy beam$^{-1}$ and the synthesis beam {\che size is} $0\as51 \times 0\as31$ (P.A. = $63.05^{\circ}$). The velocity resolution of \DCOp data is 0.085 kms$^{-1}$.

\subsection{SO ($\nu=0, J = 8_{8}-7_{7}$) data}
  In this paper, we present the results of the newly released ALMA Cycle 4 SO $\nu=0, J = 8_{8}-7_{7}$ {\che archival} data (project ID: 2016.1.01468.S, PI: Victor Magalhaes) to trace accretion shocks (See \autoref{table:1}). The observation is carried out by ALMA on March 4 in 2017 with a maximum UV baseline of $\sim$ 250 m. The data is calibrated by the ALMA calibration script, and the imaging is carried out via the CASA task CLEAN with a weighting parameter of robust 0.0. The resulting synthesized beam has an angular resolution of $1\as11 \times 0\as76$ (P.A. = $82.59^{\circ}$) with rms noise level at 20 mJy beam$^{-1}$. The largest {\ch recoverable} angular scale is around $3\as5$. The spectral resolution is 0.212 kms$^{-1}$, and the SO ($\nu=0, J = 8_{8}-7_{7}$) data have a line width larger than 8 kms$^{-1}$ resulting in {\che a} line width to channel width ratio $\sim$ 37. 



\section{{\cht Analysis}}
\label{sec:analysis}

\subsection{The comparison between Cycle 0 and Cycle 2 data} 
\label{sec:3.1}

\begin{figure}[tbh!]
\centering
\includegraphics[width=\hsize]{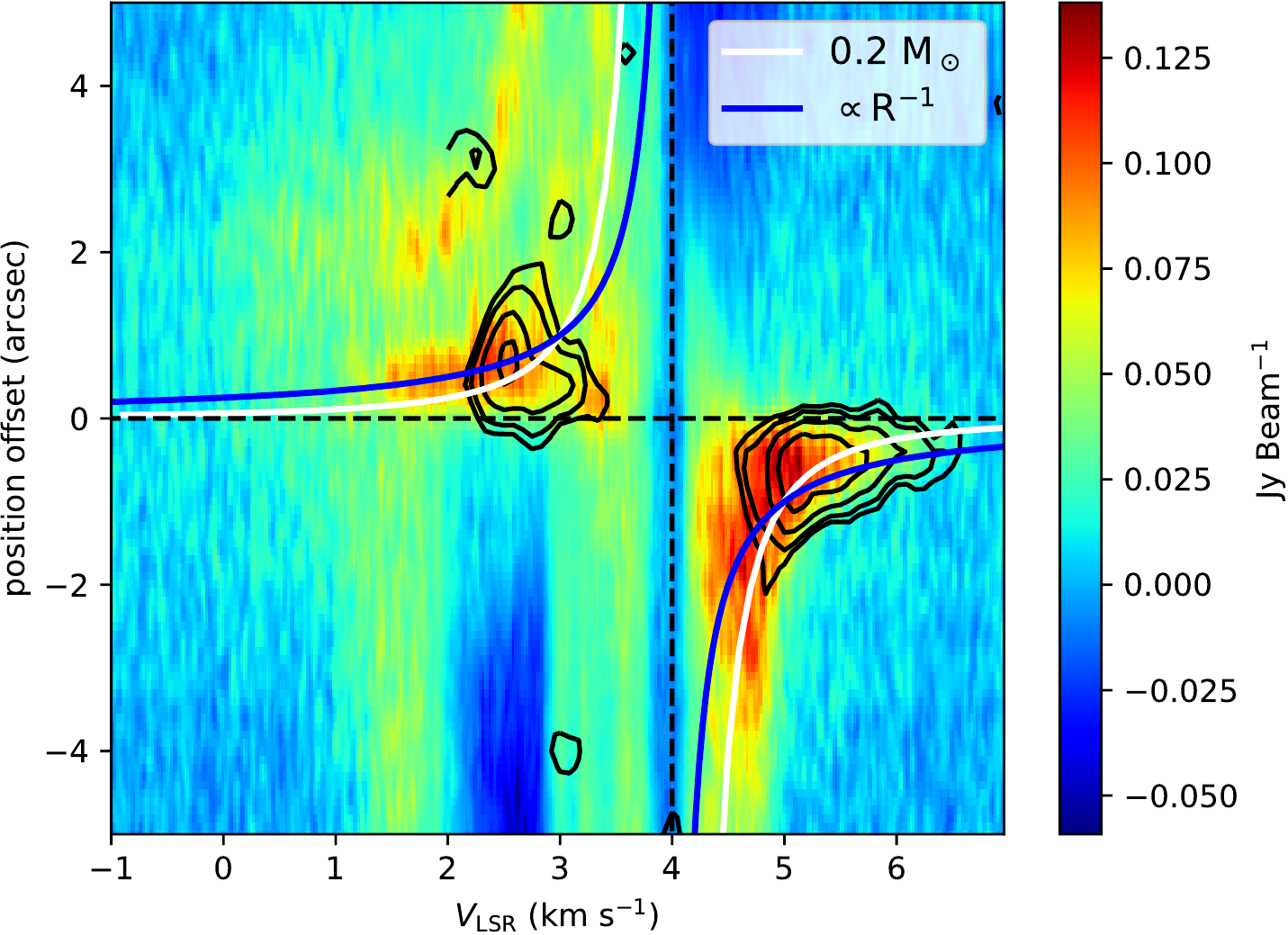}
\caption{VLA1623A C$^{18}${\cht O} J = 2-1 PV diagram {\chni for Briggs -1.0}.  The black contours marked the ALMA Cycle 0 {\che C$^{18}$O J=2--1} data, and the color background represents ALMA Cycle 2 {\che C$^{18}$O J=2--1} data. The blue line represents the {\chni infall} velocity profile with conserved specific angular momentum. The white line represents the Keplerian rotation profile with a central mass of 0.2 $M_{\odot}$. }
\label{fig3}
\end{figure}
\autoref{fig3} shows the comparison between the Cycle 0 (project ID: 2011.0.00902.S, PI: Nadia Murillo) and Cycle 2 (project ID: 2013.1.01004.S, PI: Shih-Ping Lai) ALMA data. In the Cycle 2 data shown as {\chs color background}, we identified a high-velocity blue-shifted emission around 1.5 kms$^{-1}$, and the red-shifted emission extends to 4\as0. {\che These features do not show in the Cycle 0 observations because there is not enough sampling in the short baselines.}


Cycle 0 consists of only 16 antennae with a maximum baseline of $\sim$ 400 m. The beam-size is around $0\as79\times0\as61$ with velocity resolution of 0.0833 kms$^{-1}$ \citep{2013A&A...560A.103M}. The largest recoverable angular scale is around $2\as46$. In comparison, Cycle 2 {\che observations have} increased sensitivity from the larger number of antennas and larger recoverable scale. More C$^{18}$O emission is recovered. Thus, the previous disk size and the structure of VLA1623A Keplerian disk needs to be reanalyzed.




\begin{figure*}[tbh]
\centering
\makebox[\textwidth]{\includegraphics[width=\textwidth]{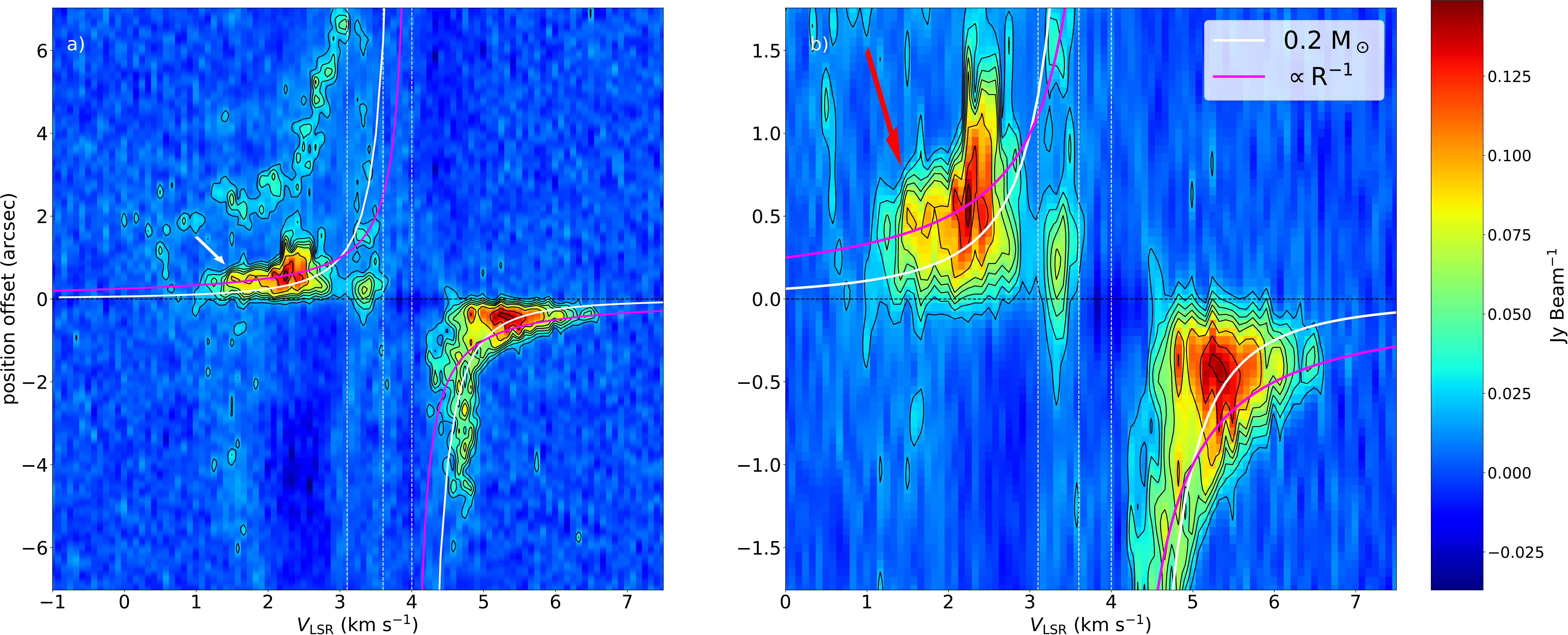}}
\caption{(a). Position-Velocity (PV) cut of {\chni the natural weighted} C$^{18}$O J = 2-1 ALMA Cycle 2 VLA1623A Keplerian disk. Centered at VLA1623A with position angle 209.82$^{\circ}$ degrees. The magenta line represents the {\chni infall} velocity profile with conserved specific angular momentum. The white line represents the Keplerian rotation profile with a central mass of 0.2 $M_{\odot}$ The white arrow marks the position of a gap between accretion flow and the Keplerian disk. {\chni The vertical white dotted lines mark the velocity of the major {\cht filtered out large scale emission}} (b). {\che The zoomed} in image of (a). Note the red arrow {\che highlights} a super-Keplerian rotation region.
}
\label{fig4}
\end{figure*}

\subsection{Position velocity diagram}
\label{sec:3.2}

To {\cht study} the gas kinematics, we create position-velocity (PV) diagrams centered at VLA1623A {\che along the red line shown in \autoref{fig2} {\cht ($\alpha$(J2000) = 16\textsuperscript{h}26\textsuperscript{m}26\fs390, $\delta$(J2000) = --24\degree24\arcmin30\as688), PA = 209.82$^\circ$) with central velocity at 4.0\,km\,s$^{-1}$}. Since the binary separation is within our beam-size, the results are the same when we shift the PV cut from one component to the other.} {\che Thus we select the center of C$^{18}$O emission for our PV diagrams}. From the rotation curves, the Keplerian rotation with a central star mass 0.2 M$_\odot$ {\chni and an inclination angle of $55^\circ$ fits} the PV data {\cht generally quite well}. This is consistent with the results of Murillo using Cycle 0 data. Thus {\cht we first set the combined mass for the VLA1623A1 and VLA1623A2 binary to be 0.2 M$_\odot$ for flat and flared disk modeling.}


In the high-velocity blue-shifted part we observed a strong emission above the Keplerian rotation white line (marked by a red arrow in \autoref{fig4}). The emission between 1.5 kms$^{-1}$ and 2.0 kms$^{-1}$ is reasonably well fitted with the {\chni infall} line (with conserved angular momentum). This feature has not been seen in the Cycle 0 observation. {\chni The scaling of the conserved angular momentum line was set such that the cross over point with the Keplerian curve is located at the centrifugal radius of the disk. The corresponding specific angular momentum is 120\,AU\,kms$^{-1}$. The detailed estimation of centrifugal radius is shown in \autoref{sec:appendix1}.}   



Is this {\che high-velocity blue-shifted component} part of the Keplerian disk? Is it due to a flared Keplerian disk with projection effects? Or is it accretion flows or other large structures in the line of sight? To determine the nature of this super-Keplerian rotation component we will conduct an analysis to identify {\cht large scale cloud emission}, accretion flows, large scale structures, and the flared Keplerian disk.

\subsection{{\cht Large scale emission around VLA1623}}
\label{sec:4.1}

\begin{figure}
\centering
\includegraphics[width=\hsize]{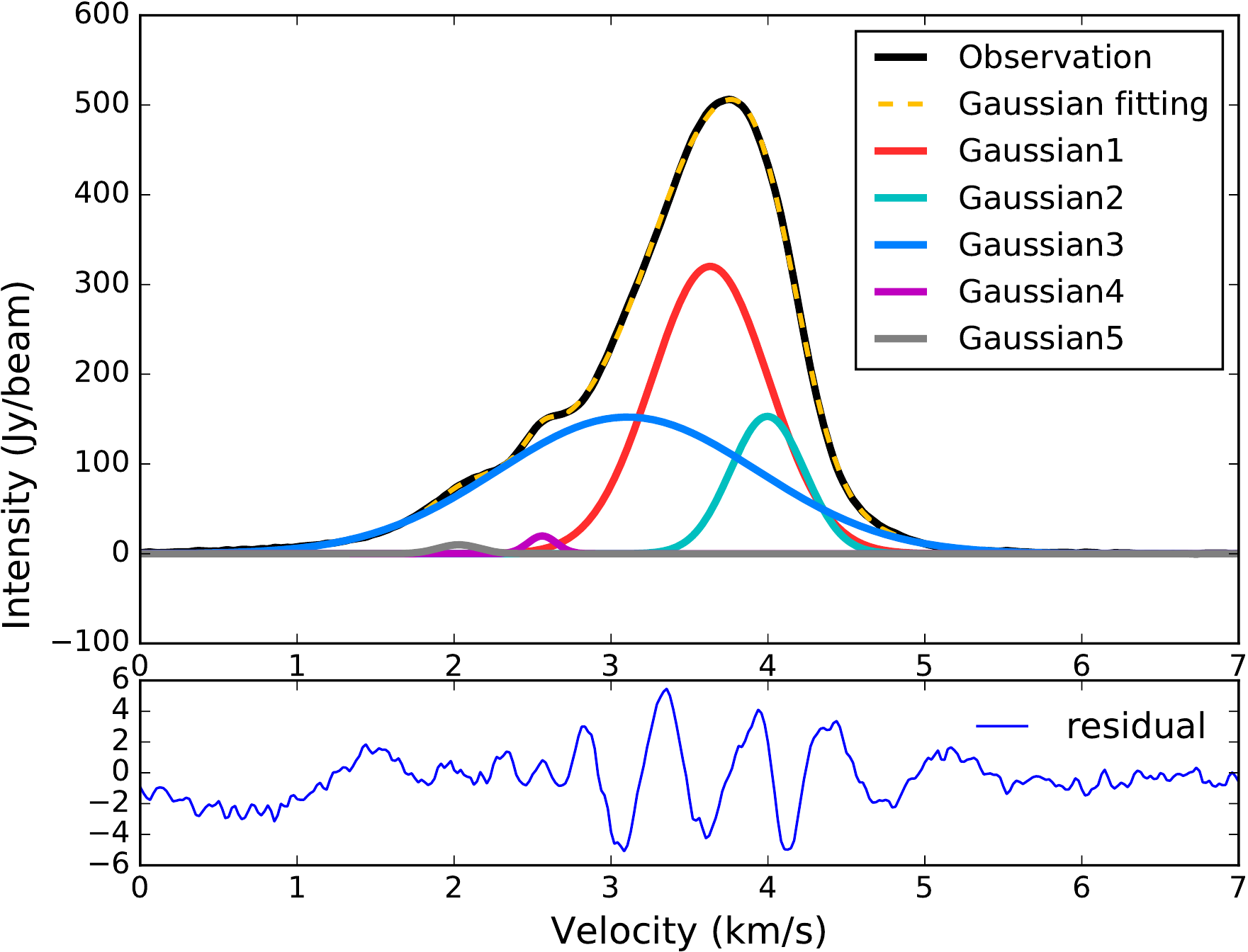}
\caption{Gaussian fit of a total power spectrum of VLA1623. The black line represents the observational data for the Total Power Array, and the color lines represent the Gaussian fitting result.}
\label{fig5}
\end{figure}

\begin{table}
\setlength{\tabcolsep}{4.6pt} 
\caption{Total Power Spectrum fitting results ({\cht Large-scale emission})}             
\label{table:2}      
\centering                          
\begin{tabular}{c c c c}        
\hline\hline                 
Number & Velocity  & Amplitude  & Widths   \\    
&(kms$^{-1}$)&(Jy)&(kms$^{-1}$)\\
\hline                        
   
   1 & 3.632 & 320.19   & 0.2782 \\ 
   2 & 3.996 & 153.11   & 0.1117 \\      
   3 & 3.104 & 152.29   & 1.3790 \\
   4 & 2.561 & 19.87   & 0.0164 \\
   5 & 2.029 & 10.04   & 0.0348 \\
   
     & $\sigma=0.353 (Jy)$ & $\chi^{2}=9401.49$ & $\chi^{2}/dof=29.29$   \\ 
\hline                                   
\end{tabular}
\end{table}

The total power spectrum data is used to determine the strength and velocities of the gas {corresponding to the large-scale emission around the } VLA1623. We did not combine the Total Power data with the 12m+7m array data in order to separate the compact disk emission from the large-scale cloud emission. In \autoref{fig5}, five Gaussian functions are used to fit the total power spectrum with the best fit result shown in \autoref{table:2}. {\chni The main component identified from the Total Power Array (3.632 km\,s$^{-1}$) is closed to the system velocity and has intensity significantly larger than the other extended emission. We therefore associate the maximum component at velocity 3.632 km\,s$^{-1}$ as} the {\cht cloud emission} of the VLA1623 system.  In \autoref{fig3} at the system velocity 4.0 kms$^{-1}$ the C$^{18}$O suffers a huge absorption and this is consistent with our Gaussian fitted {\cht extended emission} component 2 shown in \autoref{fig5}. {\chni The coincident of vertical gaps in the PV diagrams and the Total Power components suggest these vertical gaps are resulted from the spatial filtering of extended emission. In all the C$^{18}$O PV diagrams, we combined both the 12m array and 7m array to have maximum recoverable scale up to $\sim$23\as3 ($\sim$2760\,AU). The disk emission and accretion flows are well covered in this range. Total Power Array traces extended emission at scales from  $\sim$29\as7 ($\sim$3350\,AU) to $\sim$427\as9 ($\sim$51350\,AU). Any emission only observed in the Total Power Array is unlikely to be originated from the compact disk or accretion flows.}  

{\chni For} other minor components located at velocities 3.104, 2.029, and 2.561\,kms$^{-1}$, their physical origins and whether or not they are part of the VLA1623 system are unclear. Furthermore, all the {\cht filtered large scale emission} in the PV diagrams including the central envelope are located in the blue-shifted region which is consistent with the preliminary estimations done by \cite{2013A&A...560A.103M}. 

{\che The {\cht Total Power data (large scale emission)} is not included in the flat and flared disk models (Section \ref{sec:4.3} and \ref{sec:4.4}).} Since {\cht large scale emission corresponds to scales much larger than the disk or accretion flows, they are unlikely originate from the disk emission}, and the low spatial resolution of data would {\cht bias the modeling}. The purpose of this {\cht Total Power data fitting} is to identify the velocity of {\cht large scale emission}. When comparing the data with models, we will {\chn avoid these components.}

\subsection{Using Dendrograms to identify Accretion flows and set {\che constraint on} disk size}
\label{sec:4.2}

\begin{figure*}[tbh]
\centering
\makebox[\textwidth]{\includegraphics[width=\textwidth]{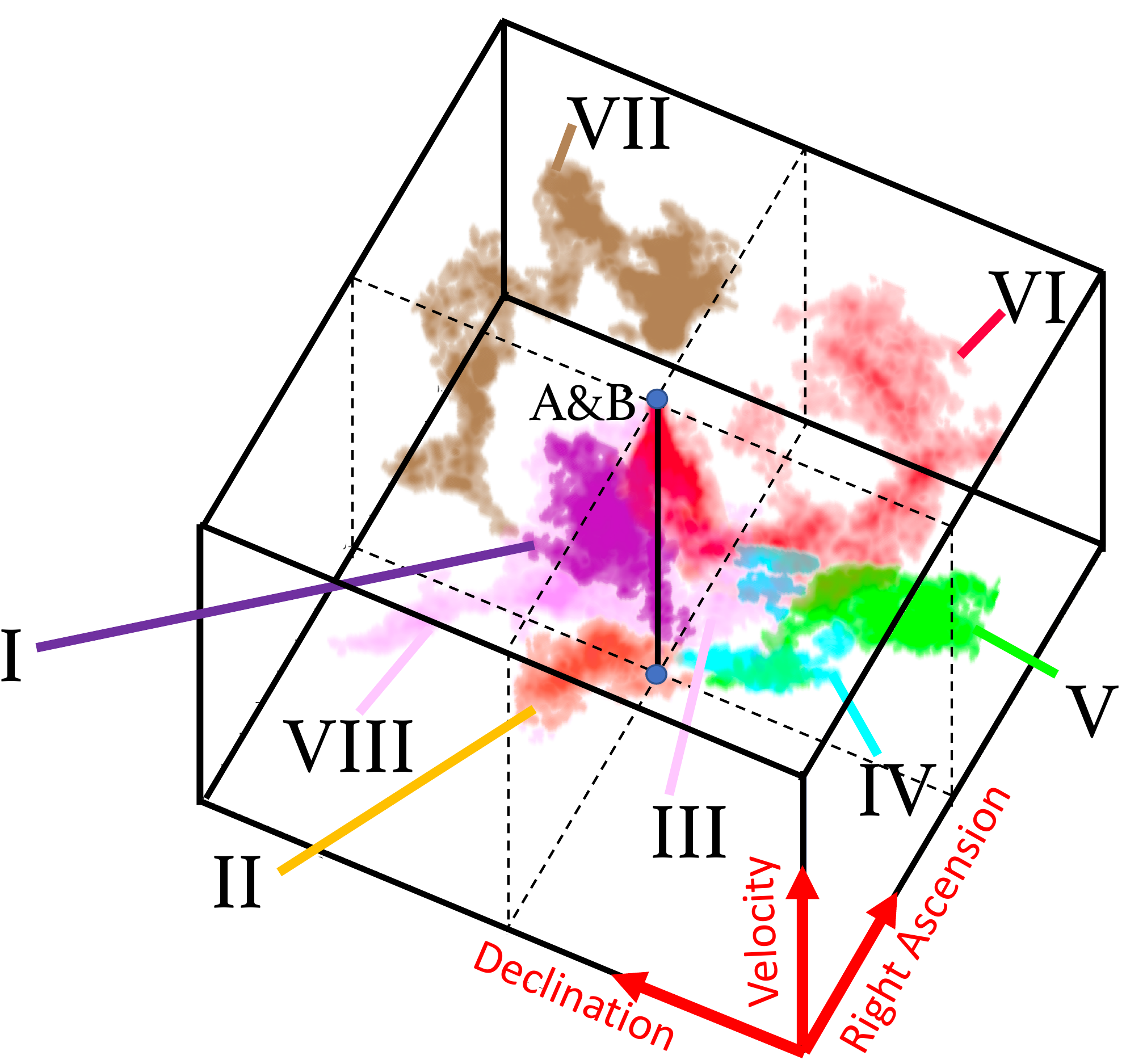}}
\caption{3D Dendrogram around VLA1623 system. The C$^{18}$O large structures are labeled with Roman numbers. The {\che central black {\cht solid} line} marked the position of VLA1623A and VLA1623B in the PPV cube. {\cht VLA1623A and VLA1623B are very close to each other with separation within the blue dot.} {\che The 3D image is made by using GLUE visualization software \citep{2014ascl.soft02002B}.} For the detailed analysis of dendrograms please see Cheong et al. {\chni (2019; submitted)}. 
}
\label{fig6}
\end{figure*}

\begin{figure*}[tbh]
\centering
\makebox[\textwidth]{\includegraphics[width=\textwidth]{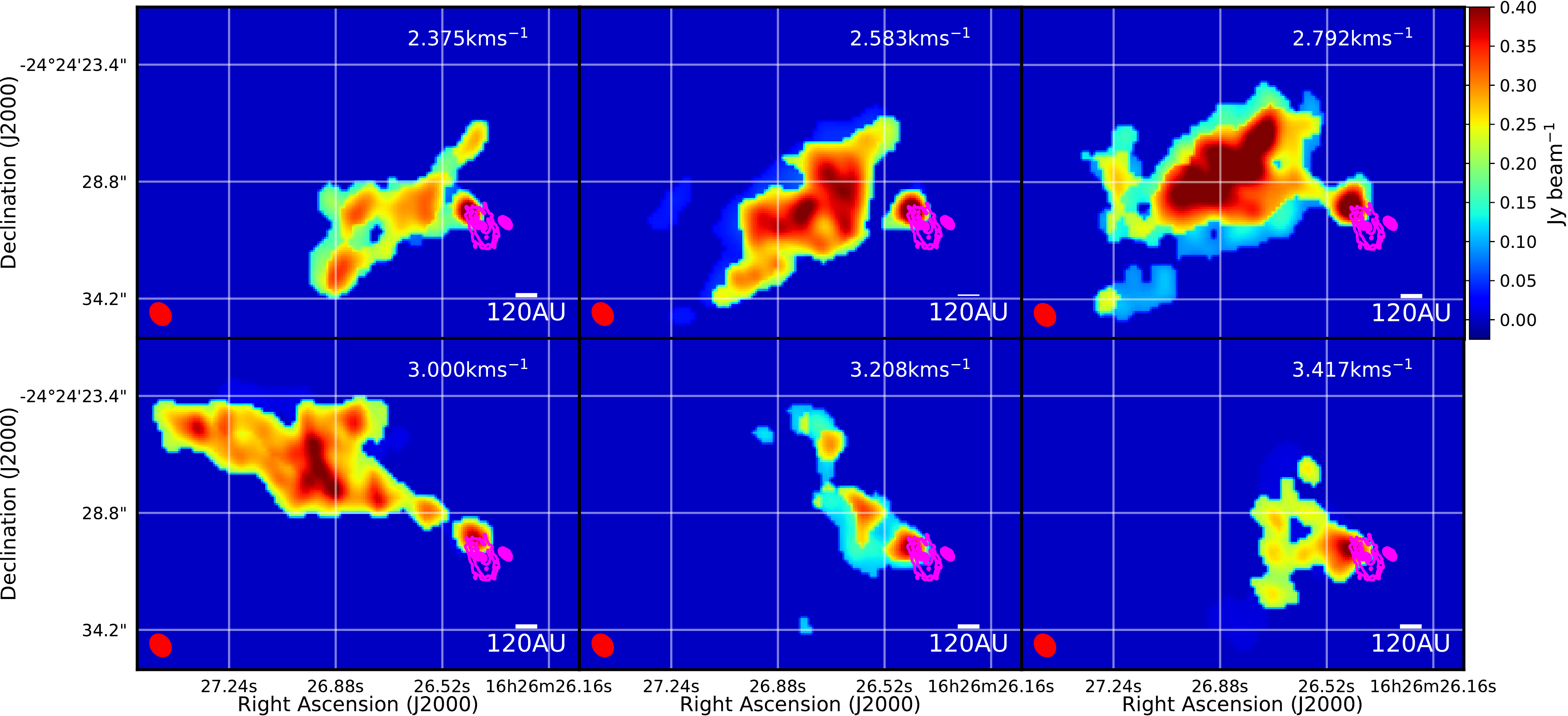}}
\caption{VLA1623A and {\chs blue-shifted} accretion flow {\chs I} channel maps.  The color represents the C$^{18}$O J = 2-1 emission {\chni (Briggs -1.5)} identified by the dendrogram algorithm. The dendrogram corresponds to the Purple component (accretion flows and disk) in Cheong et al. (in prep.)'s paper. The magenta contours are 0.88 mm continuum data. The contours are in steps of 3\,$\sigma$, 5\,$\sigma$, 10\,$\sigma$, 20\,$\sigma$, 40\,$\sigma$, 80\,$\sigma$ , where $\sigma = 5\times 10^{-4}$ Jy beam$^{-1}$). 
}
\label{fig7}
\end{figure*}

{\chn A Class 0 protostar is {\che actively accreting} and is still deeply embedded in the {\che envelope} \citep{2015MNRAS.446.2776S}. With infalling streamers feeding material from the envelope onto the circumbinary disk and outflows, it is challenging to identify disks around Class 0 sources.} In order to determine the disk size in {\che a Class 0} source, we need to identify outflows, envelope, accretion flows and other large scale structures. {\che Outflows have been observed in $^{12}$CO by }  \citep{1990A&A...236..180A,1995MNRAS.277..193D,1997ApJ...479L..63Y}. {\chel The envelope}, {\che which is located at 3.6 km\,s$^{-1}$ fitted by the total power spectrum,} can be separated out from the rest of the components in velocity domain. 

{\chni The} C$^{18}$O traces both the rotational disk and accretion flows around it. {\chni The dendrogram} algorithm is used to identify the connected structures in the {\chni position-position-velocity} (PPV) space {\chs (Cheong et al. {\chni 2019; submitted}; see \autoref{fig6}). The algorithm identifies in total {\cht 8} major {\che branches} (local maximum) {\chs labeled in \autoref{fig6}}. In additional to the 6 major branches found in {\chni Cheong et al. (2019; submitted)}, we found another {\cht 2 large structures}, blue-shifted III {\cht and VIII} component, connected to the VLA1623A circumbinary disk. In the following sections, we use SO as a shock tracer to identify the interactions between accretion flows and the circumbinary disk (See \autoref{table:3}, \autoref{fig19}).

Besides the blue-shifted III {\cht and VIII} component, {\chs Cheong et al. ({\chni 2019; submitted}) further {\chn compared the data with} the CMU model \citep{1976ApJ...210..377U,1981Icar...48..353C}, a rotating collapse model with conserved specific angular momentum, and found that the red-shifted {\che VI} component and blue-shifted I component are accretion flows connected to the central Keplerian disk ({\che channel} maps shown in \autoref{fig7} \& \autoref{fig8}). }

\autoref{fig7} {\chs shows }the channel maps of the {\chs blue-shifted I} component {\chs identified by the dendrogram.} The 0.88 mm continuum data is shown as the magenta contours which {\che mark} the location of the VLA1623A circumbinary disk and VLA1623B. Inside the magenta contours, the C$^{18}$O emission shows the {\chs blue-shifted I} accretion flows are faintly connected to the central disk. The drop of C$^{18}$O {\chs intensity between the disk and large structure I} indicates a clear gap around 120 AU exists between the large scale structure and the Keplerian disk. The Keplerian disk is constrained to roughly 180 AU from the {\chs blue-shifted} channel {\chs maps}. 

\autoref{fig8} shows the channel maps of the {\chs red-shifted accretion flow VI} connected to the central disk. In the velocity channels between 4.17 to 5.00 kms$^{-1}$, the accretion flow is well mixed with the disk and the C$^{18}$O emission {\che extends} to 6\as{\chf 0} ({\chf $\sim$} 500 AU). Only in the high-velocity channels (\,$>$ 5.1 kms$^{-1}$), the C$^{18}$O emission traces the Keplerian disk without any contamination from the accretion flow. High-velocity channels are not contaminated by accretion flows, but they only provide the information in the inner region in the Keplerian disk. Thus the disk size can't be determined by the red-shifted channels. 

From the rotation curves in \autoref{fig4}, we deduce the motion of the disk is {\che Keplerian}. And hence without loss of generality, we can assume the disk is axially symmetric and use the {\chs blue-shifted} side to constrain the disk size. {\chs We select the continuum level such that the edge of the disk matches the boundary of the gap in \autoref{fig7}.} Therefore, the circumbinary disk size is constrained to be 180\,AU. Furthermore, the extended emission between 4\,kms$^{-1}$ and 5\,kms$^{-1}$ in \autoref{fig3} is {\che likely a mixture of accretion flow and disk components. We will explore this more in the following sections.}  {\che More} technical details of dendrogram data preparation and accretion flows modeling can be found in Cheong et al. ({\chni 2019; submitted})'s paper. 

\begin{figure*}[tbh]
\centering
\makebox[\textwidth]{\includegraphics[width=\textwidth]{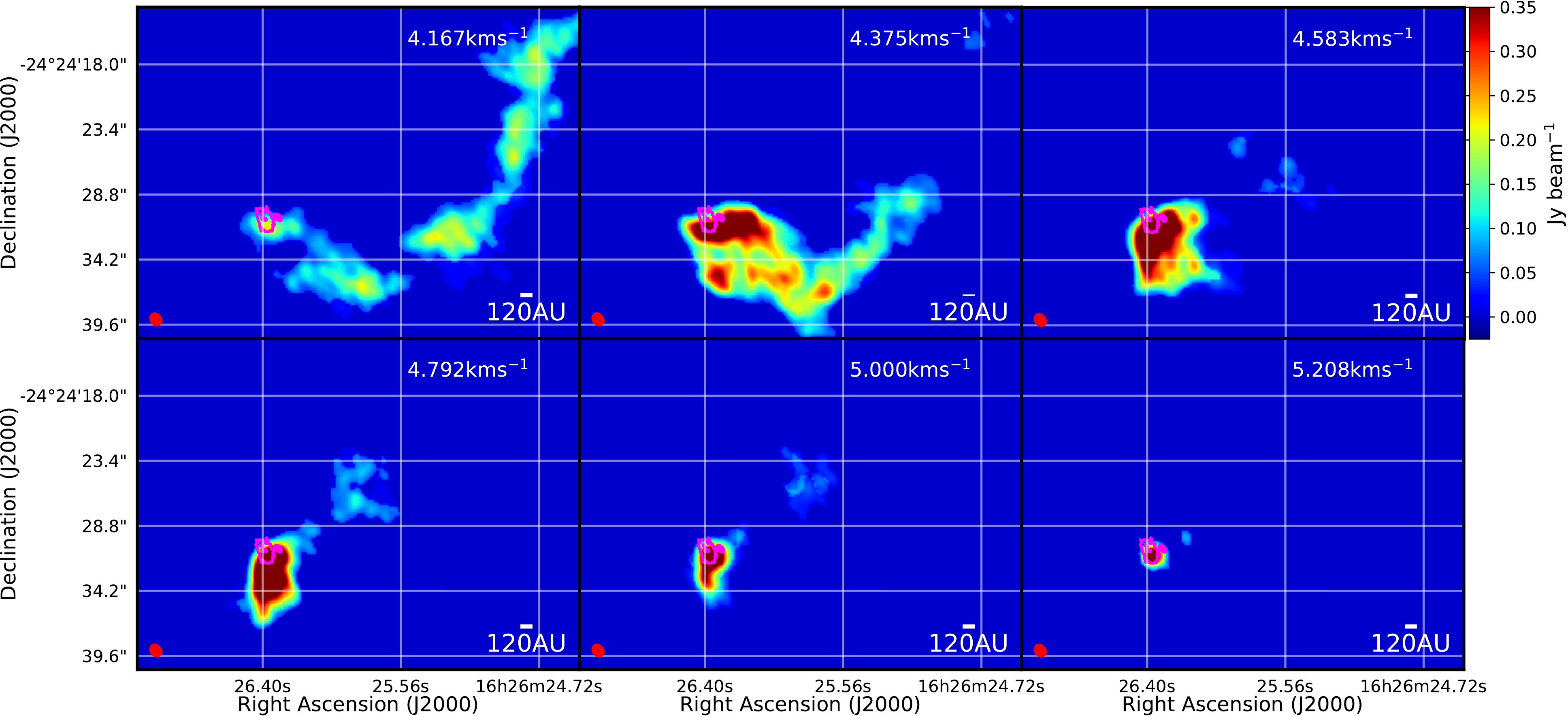}}
\caption{VLA1623A and {\chs red-shifted} accretion flow {\che VI} channel maps. The color represents the C$^{18}$O J = 2-1 {\chni (Briggs -1.5)} emission identified by the dendrogram algorithm. The dendrogram corresponds to the Red component (accretion flows and disk) in Cheong et al. (in prep.)'s paper.  The {\cht magenta} contours are 0.88 mm continuum data. The contours are in steps of 5\,$\sigma$, 40\,$\sigma$, where $\sigma = 5\times 10^{-4}$ Jy beam$^{-1}$. 
}
\label{fig8}
\end{figure*}

\subsection{Flat Disk Model}
\label{sec:4.3}

We first model the {\chs ALMA Cycle 2 C$^{18}$O J = 2--1} position-velocity (PV) diagram of the VLA1623A circumbinary disk with a Flat Keplerian disk model. The PV {\chs diagram} are cut along the red line in \autoref{fig2}. The governing equations for the velocity, {\che temperature, and column density} profiles in the flat Keplerian disk are described as the following:
   \begin{gather}
      v(R) = \sqrt[]{\frac{GM_{*}}{R}}
          \label{eq:vflat}   
    \end{gather}
    \begin{gather}
      T(R) = T_0 \times \bigg(\frac{R}{100AU}\bigg)^{-0.5}
      	   \label{eq:tflat} \\
      \Sigma(R) = \Sigma_0 \times \bigg(\frac{R}{100AU}\bigg)^{-1}
		   \label{eq:denflat}
   \end{gather}
For the column density at 100\,AU ($\Sigma_0$), we adopted the number to be {\che $6.173\times 10^{21}$\,cm$^{-2}$}, and the disk inclination angle is $55^{\circ}$ \citep{2013A&A...560A.103M}. For temperature distribution we assume the temperature power law exponent to be -0.5, and we adopted {\che a} temperature of $\sim30$\,K at 100 AU based on the DCO$^{+}$ 5-4/3-2 data \citep{2018arXiv180505205M}.


\begin{figure*}[tbh]
\centering
\makebox[\textwidth]{\includegraphics[width=\textwidth]{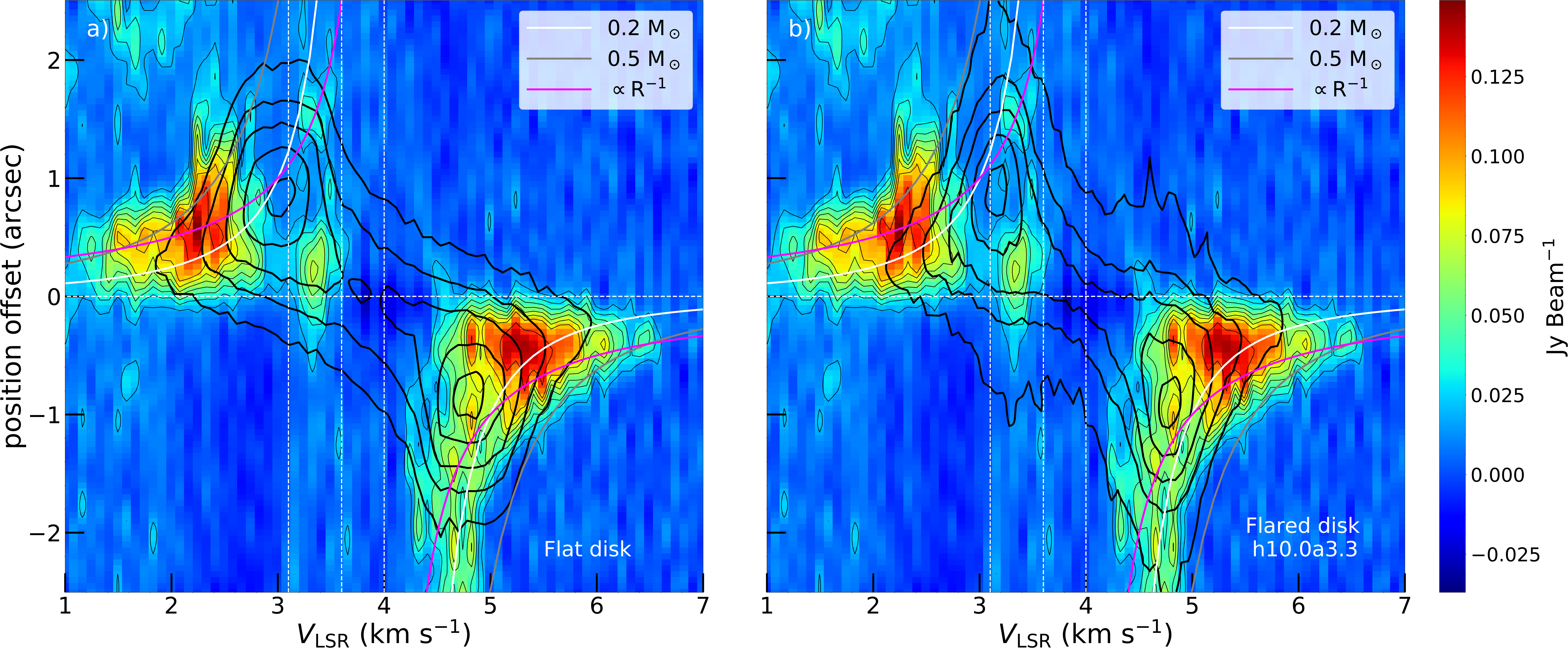}}
\caption{(a). {\chni Position velocity (PV) diagram of the VL1623A Keplerian disk overlaid with {\bf flat disk model} contours.} The color represents the PV diagram of {\chni the natural weighted} C$^{18}$O J = 2-1 emission centered at VLA1623A. The thick black contours outline the Flat Keplerian disk model in steps of 5\,$\sigma$, 10\,$\sigma$, 20\,$\sigma$, 30\,$\sigma$, 40\,$\sigma$, 45\,$\sigma$, 50\,$\sigma$, where $\sigma = 6\,$ mJy beam$^{-1}$. (b). Position-Velocity (PV) diagram of {\chni the} VLA1623A {\chni Keplerian} disk {\chni overlaid with} the {\bf best fit flared disk model contours} {\chni (vertical scale height $h_0=10.0$ and power law index $a=3.3$).} The thick black contours outline the Flared Keplerian disk model in steps of 0.2, 0.4, 0.6, 0.8, 0.95 of the maximum flux in the model. In both panels, the {\cht white, gray} and the magenta {\cht lines} represent the Keplerian rotation curve with central star mass {\cht 0.2, 0.5} $M_{\odot}$ {\chni and an inclination angle of $55^\circ$. The infall} velocity profile with conserved specific angular momentum respectively.  {\chni The vertical white dotted lines mark the velocities of the major {\cht filtered out large scale emission}.} The color {\chni background} represents the PV diagram of {\chni the natural weighted} C$^{18}$O J = 2-1 emission centered at VLA1623A and the thin black contours are plotted for 3\,$\sigma$, 5\,$\sigma$, 7\,$\sigma$, 9\,$\sigma$, 11\,$\sigma$, 13\,$\sigma$, 15\,$\sigma$ with $\sigma=9$\, mJybeam$^{-1}$.
}
\label{fig9}
\end{figure*}

A flat disk model is first generated using \autoref{eq:vflat}, \autoref{eq:tflat}, {\che and} \autoref{eq:denflat}. {\chni A simple ray tracing radiative transfer calculation scheme assuming local thermal equilibrium (LTE) with a thermal broadening of 0.2 kms$^{-1}$ is used to generate the synthetic images in position-position velocity (PPV) space.} Then the simulated disk is convolved with the ALMA telescope beam using CASA {\che SimObserve} and CASA SimAnalyze. The exact antenna setup for C34-1, C34-5 and two ACA observations {\che are} input into {\che SimObserve} to recreate the exact beam used in the observation.


{\che From} the PV diagram shown in \autoref{fig9} {\cht a}, we found the peak location of the {\che flat} disk model has a significant offset compared to the data. The {\che flat} disk model has a peak {\che located} at 0\as8 position offset while the observational data have a peak {\che located} at 0\as5 position offset in the red-shifted side. No emission was detected in the observational data corresponding to the {\che flat} disk model's blue-shifted peak, and {\che this is consistent with the {\cht filtering of the large scale emission} at 3.1\,km\,s$^{-1}$ fitted by {\cht the} total power spectrum.} 

We plot both the {\chni infall} with conserved angular momentum and Keplerian rotation curves in \autoref{fig9} {\cht a}. In the outer region of the disk, the white Keplerian line passes through the {\che flat} Keplerian disk model and the observation data on the red-shifted side for position offset within 2\as5. However, in the corresponding zoomed out PV diagram in \autoref{fig4} a, for position offset greater than 2\as5 the white Keplerian rotation line clearly deviates from the data. Thus it is very likely the C$^{18}$O long tail between 2\as5 to 5\as0 corresponds to materials {\chel at system velocity slowly} infalling towards the circumbinary disk. 


\subsection{Flared Disk Model and the {\che constraint} of VLA1623 circumbinary disk's vertical scale height}
\label{sec:4.4}

{\cht A more sophisticated 3D flared disk model is further developed to constrain the density profile and the physical structure of the circumbinary disk around VLA1623A.}
We followed the equations of Guilloteau \& Dutrey to develop a 3D flared disk model \citep{1998A&A...339..467G,2014ApJ...793....1Y}. The density, velocity and temperature profile {\che are} given as the following:
   \begin{gather}
      v(R) = {\cht \bigg({\frac{GM_{*}}{R}}\bigg)^{-v}}
          \label{eq:flaredv}       
   \end{gather}
    \begin{gather}
     T(R) = T_0 \times \bigg(\frac{R}{100AU}\bigg)^{-q}
      	   \label{eq:flaredt}\\    
      \rho(R) = \rho_0 \times \bigg(\frac{R}{100AU}\bigg)^{-a}\times exp\bigg(-\frac{z^{2}}{2h(R)^{2}}\bigg)
		   \label{eq:flaredden}
   \end{gather}
And the scale height relationship is given as:
\begin{eqnarray}
      h(R) = h_{0}\times \bigg(\frac{R}{100AU}\bigg)^{b}
          \label{eq:flaredh}       
   \end{eqnarray}
$h_{0}$ is the scale height of the circumbinary disk at 100 AU, and b is the flaring index. Assuming hydrostatic equilibrium, the scale height $b=1+v-q/2$ with $v=0.5,\, q=0.4$ for a theoretical flared Keplerian disk \citep{2011A&A...529A.105G}. The value of b is set to 1.29, using the theoretical model of flared disk  \citep{1997ApJ...490..368C}. The value of $a$ follows the  $a = p+1+v-q/2 = 1.3+p$\,, with p being the power law index of surface density $\Sigma = \Sigma_0r^{-p}$ {\chf \citep{1998A&A...339..467G}}. Considering the typical range for surface density power law $1\leq p \leq 2$ we explore the density power law index $a$ in the range between $2.0\sim 4.0$.

\begin{figure*}[tbh] 
\centering
\makebox[\textwidth]{\includegraphics[width=\textwidth ]{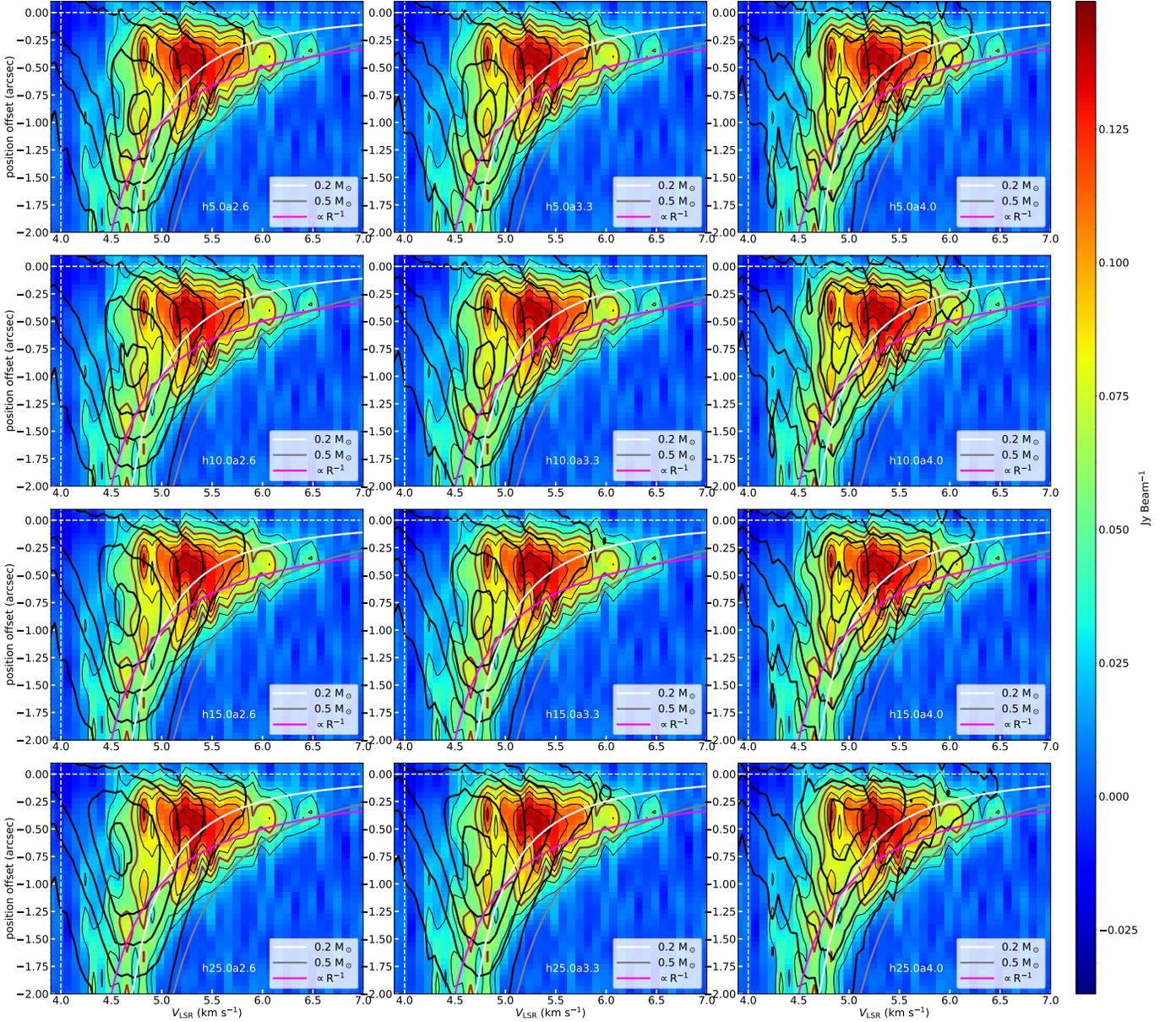}}
\caption{Position-Velocity (PV) diagram of {\chni the} VLA1623A {\chni Keplerian} disk {\chni overlaid with flared disk model contours {\cht (central binary mass: 0.2 $M_{\odot}$)} for different vertical scale height $h_{0}$ and density power law $a$.}   The color represents the PV diagram of {\chni the natural weighted} C$^{18}$O J = 2-1 emission centered at VLA1623A and the thin black contours are plotted for 3\,$\sigma$, 5\,$\sigma$, 9\,$\sigma$, 11\,$\sigma$, 13\,$\sigma$, 15\,$\sigma$ with $\sigma=9$\,mJybeam$^{-1}$. {\che The brown contour marks the 7\,$\sigma$ line used for model comparison.} The thick black contours are from the Flared Keplerian disk model. The contours are in steps of 0.2, 0.4, 0.6, 0.8, 0.95 of the maximum flux in {\chf the model}. The white {\cht and gray lines represent} the Keplerian rotation curve with central star mass 0.2 {\cht and 0.5} $M_{\odot}$ {\chni and an inclination angle of $55^\circ$}. The magenta line represents the {\chni infall} velocity profile with conserved angular momentum.
}
\label{fig10}
\end{figure*}

In our simple flared disk model we set the C$^{18}${\che O} to H$_{2}$ ratio to be $1.7\times 10^{-7}$. The resolution of the model is $500\times 500 \times 500$ with each pixel at the resolution of 0.72 AU. For the density profile, we normalized the $\rho_0$ such that the total mass of the circumbinary disk is 0.02 $M_{\odot}$ (Cheong et al. {\chni 2019; submitted}). The disk size is set to be 180 AU in radius {\che as} constrained from the C$^{18}$O data. The inclination angle is 55$^{\circ}$ and the distance is at 120 pc \citep{2008ApJ...675L..29L,2013A&A...560A.103M}. 
 

{\cht We apply the} RADMC-3D {\cht radiative transfer} code {\cht on flared disk model to create} synthetic images in 3D Cartesian geometries  {\cht with channel width set to 0.060 kms$^{-1}$} \citep{2012ascl.soft02015D}\footnote{RADMC-3D website: http://www.ita.uni-heidelberg.de/$\sim$ dullemond/software/radmc-3d/}. {\che The synthetic images are then convolved with the ALMA telescope beam using CASA {\che SimObserve} and CASA SimAnalyze.} Since we are interested in the density structure ($a$) and the vertical structure ($h_{0}$) of the flared circumbinary disk, we fixed the temperature $T_{0}$ to be 30K at 100AU {\ch based on the DCO$^{+}$ data \citep{2018arXiv180505205M}.}  

{\chs \autoref{fig10} display {\che various} flared disk models with different} density power law index $a$ and the vertical scale height parameter $h_{0}$ for the red-shifted side of the disk. Comparing the models in \autoref{fig10} and \autoref{fig21} with observation data is very challenging. The total power spectrum {\ch reveals} a huge {\cht large scale emission} at velocity 4.0 kms$^{-1}$ {\cht (systematic velocity)} and many {\cht more} on the blue-shifted side. Therefore, in order to eliminate the {\cht contamination of the large scale cloud, foreground and background emission}, {\che we only model the red-shifted part on the right side of the line corresponding to system velocity 4.0 kms$^{-1}$.} {\cht Furthermore, to} avoid the contamination from the outer accretion flows, we only compare the data and model within the disk radius ($\leq$ 1\as5). 


{\chs Isophote contours for} each simulation model are plotted in order to compare the observational data with simulation models. Overall we run the parameters $h_{0}= 1.0,\,2.5, \,5.0,\,10.0,\,15.0,\,25.0$ (AU) and $a = 2.0,\,2.3,\,2.6,\,3.0,\,3.3,\,4.0$ {\che for a total of 36 parameter combinations} {\chs to examine how the central peak location changes when the scale height $h_{0}$ and the density power law index $a$ are varied}.  

 In \autoref{fig10} as the vertical scale height $h_{0}$ increases, the model's peak (thick black contours) would shift towards the connecting bridge between blue-shifted and red-shifted region {\che (system velocity 4 kms$^{-1}$, and offsets 0.0)}. As for the density power law index $a$, when $a$ increases the PV diagram would be stretched in the direction of the Keplerian rotation line (white line). This is because when $a$ increases, the extended part of the disk would be suppressed and the intensity peak would move closer to the disk center.  

At first sight, the best fit model would be the one with density power law index $a =4.0$, and scale height $h_{0} = 25.0$. The density power law index $a =4.0$ suggests the peak is very compact in the center of the circumbinary disk VLA1623A. In the very high velocity region $v>6.0$\,kms$^{-1}$ and $v<2.0$\,kms$^{-1}$ , the observation data deviates from the {\che flared} disk model {\cht and doesn't follow the white Keplerian rotation line}. This deviation suggests that an inner compact structure exists inside the VLA1623A circumbinary disk. {\cht To avoid the contamination from the inner super-Keplerian region when constraining the circumbinary disk properties, we only model the data outside the inner region ($\sim$0\as5 (60 AU)) where the inner structures of VLA1623A lies}. 


For the model with parameters $a=4.0,\, h_{0} = 25.0$, we found the peak of the flared disk model has the same position offset ($\sim$ 0\as5) as the observation data but at a much lower velocity {\cht (4.9\,kms$^{-1}$ as compared to 5.2 kms$^{-1}$)}. The peak of the flared disk model and observational data does not overlap suggests that the circumbinary disk does not have density power law index $a$ as large as 4.0. Instead, it has a relatively {\che flatter} density power law plus a compact inner structure within 60 AU from the disk center. 

For density power law index $a$ less than 3.3, there are only minor differences when $a$ varies. The PV diagram overall is insensitive to parameter $a$. To prevent contamination of {\cht large scale emission} and internal structures, we search for disk models with parameter $a\leq 3.3$, peak locations {\che between 0\as5 $\sim$ 1\as5} in position offset, and have {\chs emission} between {\che $4.5\sim 4.9$} kms$^{-1}$. 

Close inspection of the {\che fitting reveals that} when the vertical scale height $h_{0}$ equals to 25.0 AU, the models have peaks locate at velocity $4.5\sim 4.7$ kms$^{-1}$ on the red-shifted side. Most importantly, in all the $h_{0}=25.0$ models more than half of the peak area falls outside the $7\, \sigma$ (63 mJy {\che brown line in \autoref{fig10}}) line on the red-shifted side. This significant offset shows that the $h_{0}=25.0$ does not fit the data. {\cht For vertical scale height $h_{0}=15$ AU, the peak overlap area is around 20\,\% for model $h_{0}=15,\,a=2.6$, but around 70\,\% for $h_{0}=15,\,a=3.3$. Thus it cannot be completely ruled out. For vertical scale height $h_{0}$ less than or equal to 10 AU, the overlap region is more than 50\,\%.} \textbf{\che Therefore, we can only constrain the density power-law $a$ for the VLA1623A circumbinary disk to be $ a \leq 3.3$ with the vertical scale height at 100 AU to be $h_{0} \leq 15\,$ AU.} 


 
It is important to highlight again that for all of the parameter sets, no model can perfectly fit the location of the peak in the PV diagram. Both flat and flared disk model does not produce peaks at the same location as the observational data. Changing vertical scale height can't produce high-velocity peaks in the inner region of the Keplerian disk, and adjusting density power law only stretches the contours along the Keplerian rotational line. {\cht From the flat and flared disk modeling, we have shown that simple Keplerian rotation disk models with combined binary mass of 0.2\,$M_\odot$ cannot fully explain the observational data.} 

\subsection{{\chni Limitations and Degeneracy in Modelling}}
{\chni Modeling and constructing a coherent picture from the complex data set around the Class 0 source VLA1623 is difficult, as multiple physical processes (outflows, infalls, rotation, shocks) need to be taken into account, and a wide range of parameters can be adjusted. A careful and logical reasoning is required to connect the pieces and form the overall picture. In this section, we will discuss the limitations and the logical reasoning behind breaking the degeneracy in the modelling.}

\subsubsection{{\chni Important parameters and limitations of the disk modeling}}

{\chni Since the flared disk model is used to constrain disk properties (disk scale height and density power law index), the discussion of important parameters and limitations of the disk modeling will be focused on the flared disk model. 

In total there are {\cht 10} parameters in the flared disk model: scale height, temperature profile (power law index and normalization), density profile (power law index and normalization), inner cutoff radius, disk inclination, disk size, combined binary mass, and distance. {\cht Two} parameters, density profile (power law index) and scale height, are free parameters that are explored and constrained. The other {\cht 8} parameters and their effects on the PV diagram modeling (\autoref{fig10}) are listed below:
\begin{enumerate}
\item Temperature profile (power law index):

{\cht The profile} affects the position of the peak in the disk by stretching the peak along the Keplerian rotation line in the PV diagram. This is the second largest uncertainty in the modeling. It will be discussed more in {\cht \S~3.7.2}. However, we do not expect huge deviation from the theoretical flared disk model unless other heating or cooling mechanisms are present.    

\item Temperature normalization at 100\,AU:

Affects the overall normalization of the flux. Does not change the peak position in PV diagram, therefore has no or little effect on the scale height and density power law modeling.

\item Density normalization {\cht (Disk mass)}:

Affects the overall normalization of the flux. Also has little or no effect on the peak position in the PV diagram. It is constrained by normalizing the disk mass to 0.02 M$_{\odot}$ (Cheong et al. {\chni 2019; submitted}). 

\item Disk size:
 
The gas disk size is {\cht $\sim180$\,AU} constrained by the inner edge of the gap in \autoref{fig4} and \autoref{fig7}. Disk size affects the spatial size of the model in the PV diagram. This parameter is well constrained by observation, and has uncertainty around half of the beam size ($\sim$30\,AU).

\item inner cutoff radius:

Since the disk is in Keplerian rotation, the closer to the protostar the faster it rotates. The inner cutoff radius would determine the velocity cutoff point in the PV diagram. To prevent artificial velocity cutoff at high velocity, small inner cutoff radius is chosen. In the inner region of the disk since the area decreases with smaller radius, the intensity drops rapidly at high velocity end of the PV diagram. Therefore, for small enough inner cutoff radius the velocity cuttoff would have little or no effect on the disk modeling. In the flared disk modeling, the inner cutoff radius is set to be 1\,AU (cell size 0.72\,AU). As for the flat disk model the inner cutoff radius is also 1\,AU.

\item Disk inclination

The inclination angle used in the modeling is 55$^{\circ}$  \citep{2008ApJ...675L..29L,2013A&A...560A.103M}. The PV diagram is aligned along the major axis of the disk, so the change in inclination would have no effects in the direction of position offset in the PV diagram. Inclination only affects the projection of velocity to the line of sight and stretches the model contours away or towards from the systematic velocity. Even for 5$^{\circ}$ variation of inclination angle, the velocity stretching factor is less than 7\,\%. In \autoref{fig10}, for density power law index $a<3.3$ the peak velocity difference between vertical scale height $h=15$\,AU and $h=25$\,AU is $\sim 0.2$\,kms$^{-1}$. The peak velocity of the $h=25$\,AU models are located around 0.6 to 0.7\,kms$^{-1}$ away from the systematic velocity. If the $h=25$\,AU models are stretched by 7\,\%, the peak velocity would increase at most by $\sim 0.05$\,kms$^{-1}$, which is still smaller than the $0.2$\,kms$^{-1}$ difference used to distinguish accepted and non-accepted models. Thus disk inclination in our modeling would not change the result of flared disk modeling.



\item Combined binary mass

Combined binary mass is the most important factor in the disk modeling. Not only will it significantly affect the results on flared disk modeling, it {\cht is} also {a possible explanation for the super-Keplerian rotation in the inner region of the disk. More} in-depth discussion about the degeneracy of combined binary mass would be presented in {\cht \S~4.1.}

\item Distance

In this paper, the distance of 120\,pc \citep{2013ApJ...764L..15M} is used in the modelling instead of 137.3\,pc \citep{2017ApJ...834..141O}. The physical scale would increase by a factor of 1.14 if the 137.3\,pc distance is adopted. In \autoref{fig10}, for power law index $a<3.3$, the shape of the model peaks are elongated in the direction of position offset. One major difference between vertical scale height $h=15$\,AU and vertical scale height $h=25$\,AU is that the centers of the peaks are located at different velocities. Stretching the models by 14\,\% in position offset would make it harder to differentiate between the two, but it will not change the result of the flared disk modeling.

\end{enumerate}
}

\subsubsection{{\chni Degeneracy in Temperature and Density profile}}
\label{sec:dis-Temp}

{\chni Disk intensity profile is determined by both temperature and density. In this study, for the flared disk model we assumed a fixed temperature profile with a power law index of -0.4 based on the theoretical flared Keplerian disk model \citep{2011A&A...529A.105G}. As for the normalization we fixed the temperature to 30\,K at 100\,AU based on the previous DCO$^{+}$ modeling \citep{2018arXiv180505205M}. 

Fixing the temperature profile greatly reduces the free parameters in the modelling. It is important to note that the temperature profile is based on a simple theoretical model and in reality the actual temperature profile might deviate from this fixed profile. If the power law index is varied with the overall temperature normalization fixed at 100 AU, you would expect the peak to move inward or outward depending on the power law index. In other words, the peak will move along the Keplerian rotation line in the PV diagram. This will affect the density and scale height modeling, making it even more difficult to break the degeneracy, but the model still won't be able to reproduce the {\cht observed} peaks which clearly deviates from the Keplerian rotation. Thus it {\cht is not a good explanation for the super-Keplerian motion}.

It will, however, affect the result of scale height modeling. For a larger disk scale height, the peak will move closer to the position offset 0.0 (center of the disk) and towards the system velocity in the PV diagram. Adjusting the temperature power law index would further stretches the peak along the Keplerian line. The two completing effects combined would make it very difficult to break the degeneracy. Even so, we do not expect the temperature profile to deviate too much from the theoretical flared disk model unless additional heating or cooling processes are involved. 

The better and more accurate way to break the temperature and density degeneracy is to use two C$^{18}$O transitions to obtain the temperature profile and constrain it directly from observation instead of using the simple theoretical model. Future observation is needed to more accurately constrain the temperature profile of the VLA1623A circumbinary disk. }



\subsection{{\cht Accretion shocks around the circumbinary disk}}
\label{sec:4.5}

\begin{figure*}[tbh]
\centering
\makebox[\textwidth]{\includegraphics[width=\textwidth]{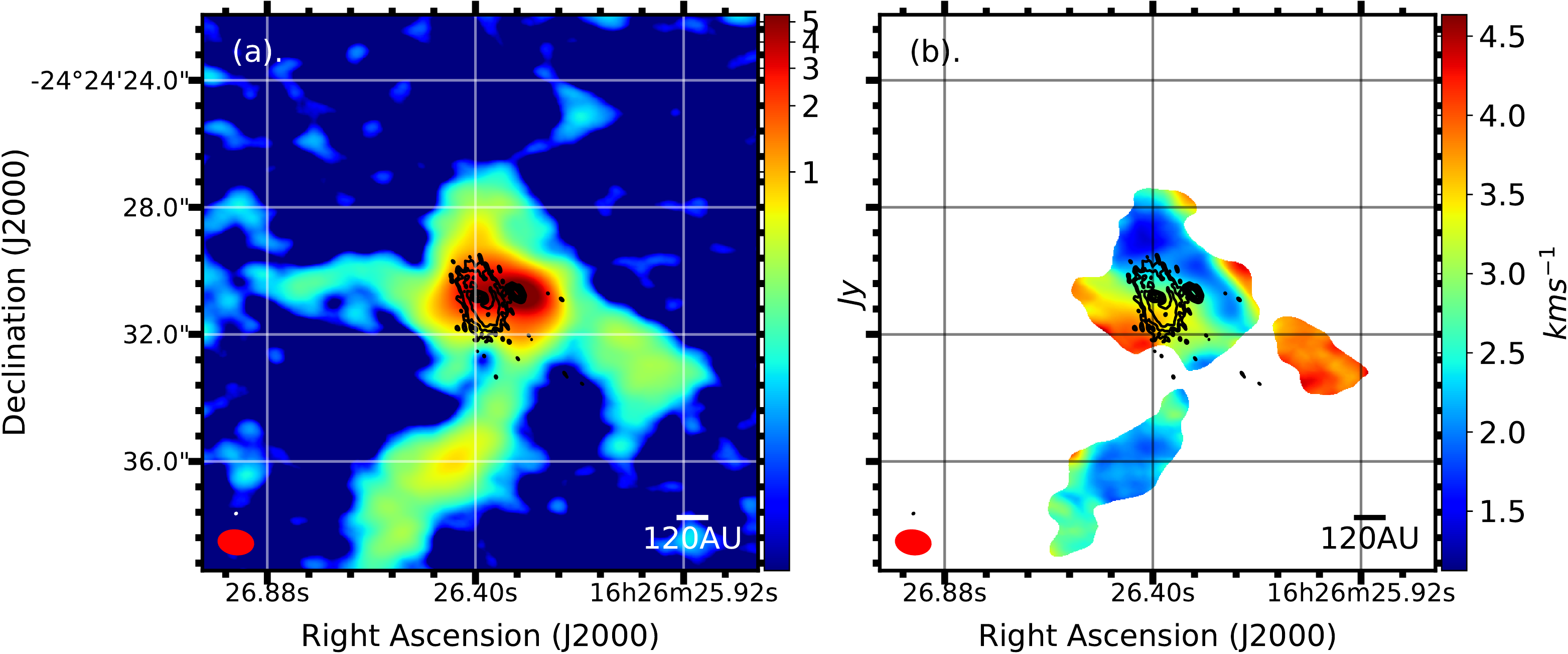}}
\caption{(a). SO intensity integrated map (Moment 0).  The color represents the ALMA Cycle 4 SO $\nu=0, J = 8_{8}-7_{7}$ data. The black contours are 0.88 mm continuum data. The contours are in steps of 3\,$\sigma$, 5\,$\sigma$, 10\,$\sigma$, 20\,$\sigma$, 40\,$\sigma$, 80\,$\sigma$ , where $\sigma = 5\times 10^{-4}$ Jy beam$^{-1}$. (b). SO mean velocity map (Moment 1).  
}
\label{fig11}
\end{figure*}

\begin{figure*}[tbh]
\centering
\makebox[\textwidth]{\includegraphics[width=\textwidth]{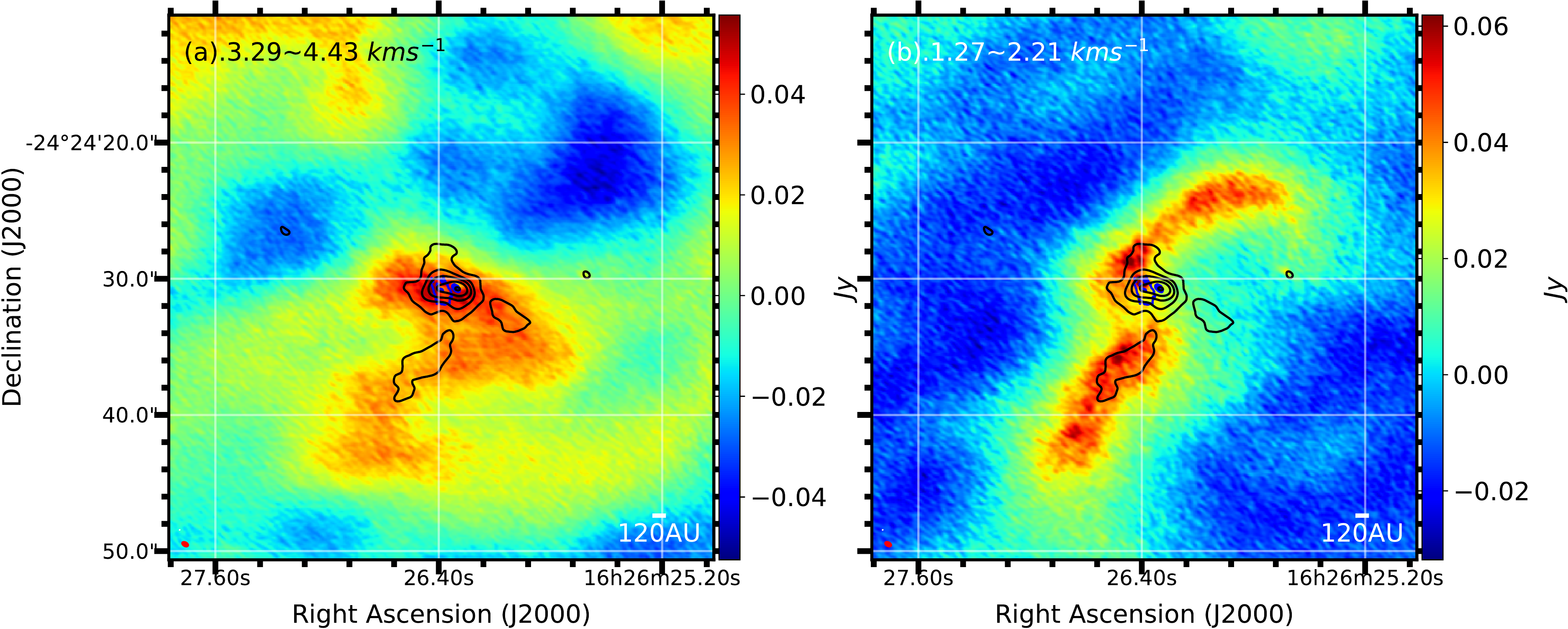}}
\caption{(a). C$^{18}$O J = 2-1 intensity integrated map (Moment 0) between 3.29 kms$^{-1}$ and 4.43 kms$^{-1}$ {\chni for Briggs -1.0 weighted data}. (b). C$^{18}$O J = 2-1 intensity integrated map (Moment 0) between 1.27 kms$^{-1}$ and 2.21 kms$^{-1}$ {\chni for Briggs -1.0 weighted data}.  The color in both panels represent the C$^{18}$O data. The black contours are ALMA Cycle 4 SO $\nu=0, J = 8_{8}-7_{7}$ data. The contours are in steps of 0.025, 0.05, 0.1, 0.2, 0.4, 0.8 of the maximum intensity which is $ 8.5$ Jy beam$^{-1}$. The central blue contours are 0.88 mm continuum data in steps of 5$\sigma$, 40$\sigma$, where $\sigma = 5\times 10^{-4}$ Jy beam$^{-1}$. The continuum data marked the position of the VLA1623A circumbinary disk and VLA1623B.
}
\label{fig12}
\end{figure*}


{\ch SO with sublimation temperature of 50\,K is generally attached to dust grains. Observation of SO emission indicates {\che that it comes from} collisions or shocks that give enough energy to free SO into the gas phase.} Previously SO emission has been used to trace shock fronts in another similar Class 0 disk system, L1527 \citep{2014Natur.507...78S}. {\cht In this section, we use SO as a shock tracer to understand the interactions between the accretion flows and the circumbinary disk.}

\autoref{fig11} a shows the SO $\nu=0, J = 8_{8}-7_{7}$ Moment 0 map. {\chf A strong enhancement of SO molecules near the circumbinary disk and VLA1623B is apparent in the {\cht Moment} 0 map and this indicates strong accretion shocks {\cht are} created around VLA1623A circumbinary disk and VLA1623B.} The SO emission shows 4 stream-like structures with two main streams to the {\chni west} and south. The South stream has the strongest emission in all the streams and is connected to VLA1623B. The {\chni West} stream corresponds to the red-shifted accretion flow {\che VI} identified by the dendrogram analysis using C$^{18}$O, and the {\chni East} stream corresponds to the blue-shifted accretion flow {\chs I} in \autoref{fig6} ({\che Red and Purple Accretion flow respectively} in Cheong et al. ({\chni 2019; submitted})). 

\begin{figure}[h]
\centering
\includegraphics[width=\hsize]{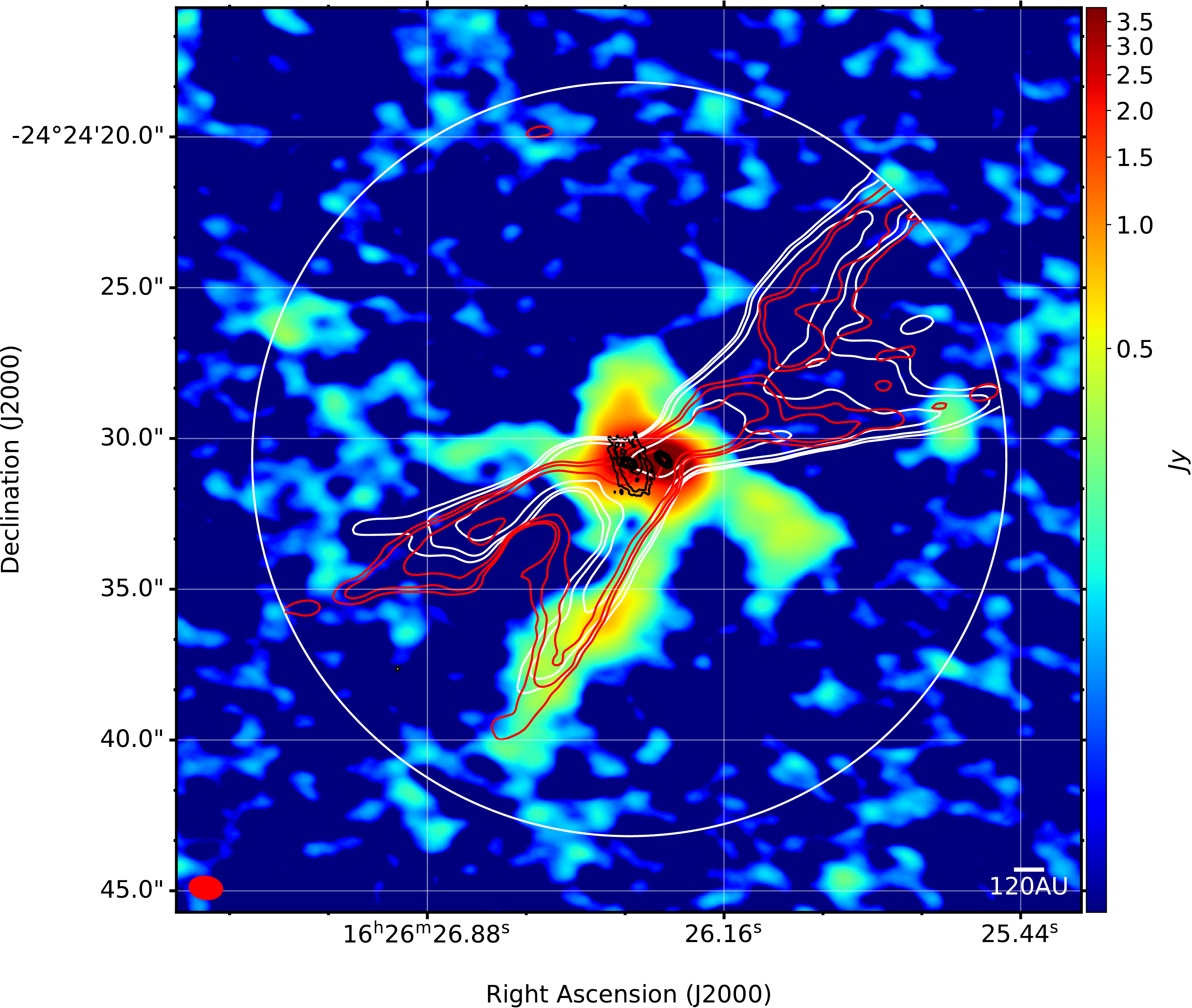}
\caption{{\chni Intensity map of ALMA Cycle 4 SO ($\nu=0, J = 8_{8}-7_{7}$, color), ALMA Cycle 6 CO (J = 3-2, red and white contours), and 0.88\,mm continuum (black contours). The CO contours are in steps of 30\,$\sigma$, 50\,$\sigma$, 100\,$\sigma$, 300\,$\sigma$, where $\sigma =$ 11\,mJy beam$^{-1}$ for CO. The red contours are red-shifted integrated CO emission between 8.80\,kms$^{-1}$ to 17.1\,kms$^{-1}$, and the white contours represent blue-shifted integrated CO emission between 5.20\,kms$^{-1}$ to 8.59\, kms$^{-1}$. The red circle marks the beam of the SO data, and the white circle represents the field of view for Band 7 CO observation.}}

\label{fig13}
\end{figure}

Both the South and {\chni West} stream connect to VLA1623B, {\che and} a strong emission of SO is {\ch detected} on VLA16232B. The SO peaks at VLA1623B and the stream morphology suggests that on VLA1623B there exists violent shocks caused by the collision between B and the outer red-shifted accretion flow VI. As for the North and {\chni East} streams, they connect to VLA1623A circumbinary disk on the map in the plane of sky. The North stream is much shorter than the South stream and this indicates the {\chni collision} is much closer toward the VLA1623A and VLA1623B from the north. In the South stream, the peak locates around 7\as0 away south from the VLA1623A binaries. 

\autoref{fig11} b shows the mean velocity map (moment 1) of SO emission around VLA1623A circumbinary disk and VLA1623B. {\che Notice} that the system velocity of VLA1623A circumbinary disk is 4 kms$^{-1}$ and nearly all of the SO in Moment 1 map does not have a velocity greater than 4.4 kms$^{-1}$. {\che The lack of red-shifted emissions in the North SO stream shows it corresponds to} the blue-shifted accretion flow {\cht III (\autoref{fig6}) flow} toward the circumbinary disk. The extended high velocity SO emission in the North stream is caused by the accretion shocks due to the collision between the circumbinary disk and accretion flow.


As for the South SO stream, it lies in the position corresponds to the {\chs red-shifted} accretion flow {\che VI} (Fig.\,\ref{fig8}). However, the accretion flow {\che VI} traced by C$^{18}$O is observed to be red-shifted and moving away from observer {\chs while the} SO South stream is blue-shifted and moving in the opposite direction. For comparison between C$^{18}$O and SO data, we plot the C$^{18}$O J = 2-1 intensity integrated map (Moment 0) between 3.29 kms$^{-1} \sim$ 4.43 kms$^{-1}$ in \autoref{fig12} a and between 1.27 kms$^{-1} \sim$ 2.21 kms$^{-1}$ in \autoref{fig12} b with SO data display as black contours in both figures. 

\begin{figure*}[tbh]
\centering
\makebox[\textwidth]{\includegraphics[width=\textwidth]{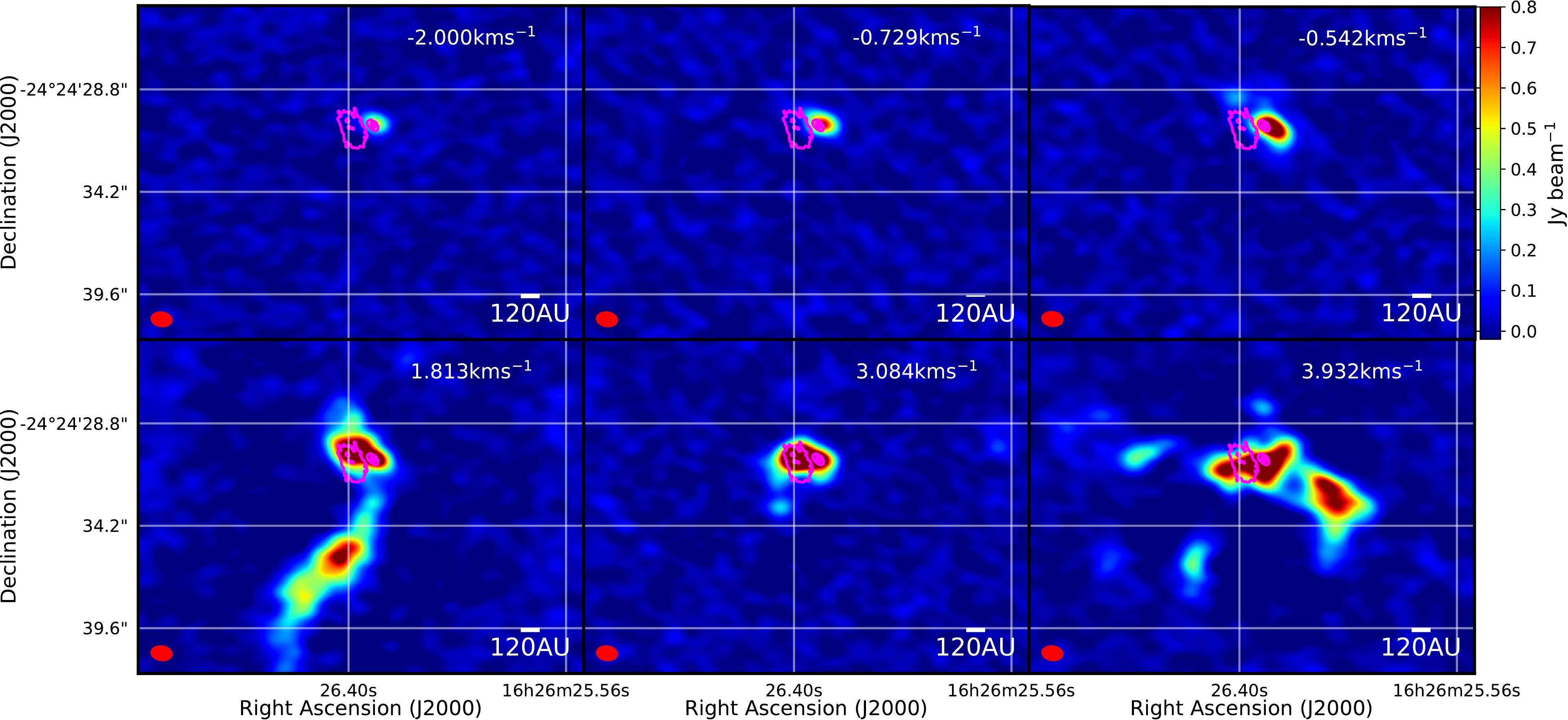}}
\caption{SO channel maps.  The color represents the ALMA Cycle 4 SO $\nu=0, J = 8_{8}-7_{7}$ data. The black contours are 0.88 mm continuum data in steps of 3\,$\sigma$, 5\,$\sigma$, 40\,$\sigma$, 80\,$\sigma$ , where $\sigma = 5\times 10^{-4}$ Jy beam$^{-1}$. 
}
\label{fig14}
\end{figure*}

In \autoref{fig12} a, the C$^{18}$O emission which corresponds to the materials at low blue-shifted velocity and at rest is much more extended than the Southern {\chs SO stream.} C$^{18}$O at rest are mostly distributed on the south side of the circumbinary disk and VLA1623B. Materials are piled up in the south, and this wall-like structure would be discussed in the next section.

One of the most prominent features in \autoref{fig12} b is the Northern and Southern arm. {\chs The Northern C$^{18}$O arm corresponds to the accretion flows III (See \autoref{fig6}).} The Southern C$^{18}$O arm has the same velocity range as the SO south stream and it overlays in the line of sight perfectly. From the dendrogram analysis carried out by Cheong et al. ({\chni 2019; submitted}), the Southern arm corresponds to the {\chs Structure II in \autoref{fig6}}. Cheong et al. ({\chni 2019; submitted}) further carried out the CMU model analysis \citep{1976ApJ...210..377U,1981Icar...48..353C}, a rotating collapse model with conserved specific angular momentum, and found the {\chs blue-shifted} component {\chs II} does not match the CMU model. This indicates the materials {\chni in} the South stream do not follow the {\chni infalling} parabolic trajectories. From the dendrogram analysis {\cht and} CMU fitting in Cheong et al. ({\chni 2019; submitted}) we {\ch concluded} that the South SO stream of  (\autoref{fig11} a \& b) corresponds to the {\chni materials} with {\chni non-conserved} specific angular momentum, {\che possibly affected by outflows from the protostellar sources.}  

{\chni To confirm this interpretation, we plot the CO outflows on top of the SO shock emission in \autoref{fig13}. The black contours represent the 0.88\,mm continuum data.  The center of the bipolar outflow coincides with VLA1623B, hinting VLA1623B might be the origin of the outflow. The Northern, Eastern and Western SO streams are further away from the outflow direction, making them more likely tracing accretion shocks from the accretion flows rather than the shock fronts of the outflows. As for the Southern SO stream, it overlays perfectly with the CO outflow suggesting the SO South stream is tracing the collision between outflow and outflow cavity walls.} {\cht The distribution of shocks in both position and velocity space are further shown as SO channel maps in \autoref{fig14}.}

\begin{figure}[tbh]
\centering
\includegraphics[width=\hsize]{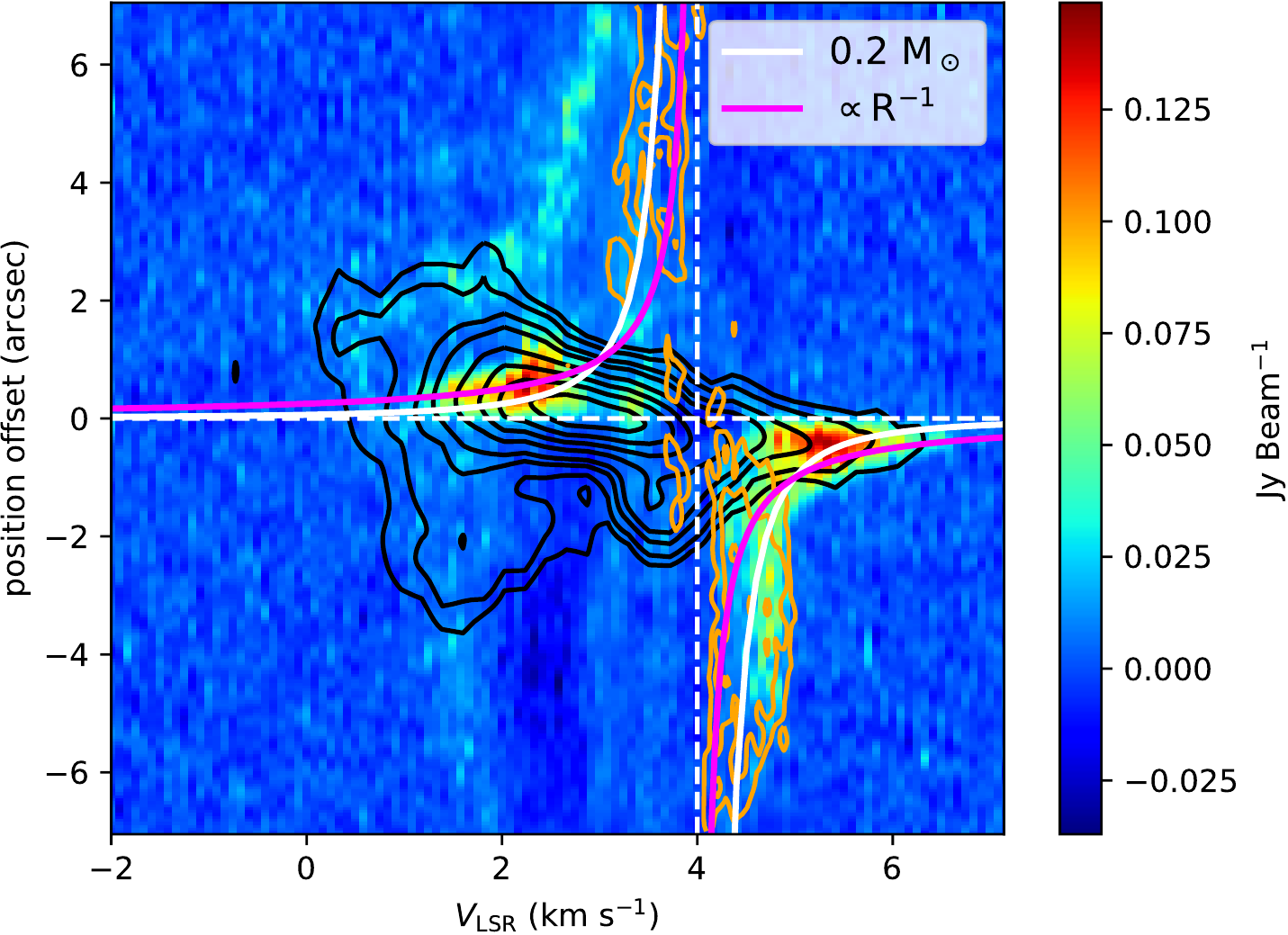}
\caption{ALMA Cycle 4 SO ($\nu=0,  J = 8_{8}-7_{7}$, black contours), ALMA Cycle 2 C$^{18}$O {\chni with natural weighting} (J = 2-1,  color), and DCO+ (J = 3-2, orange) position velocity (PV) diagram on {\bf VLA1623A}. The contours are in steps of 5\,$\sigma$, 10\,$\sigma$, 20\,$\sigma$, 30\,$\sigma$, 50\,$\sigma$, 70\,$\sigma$, 90\,$\sigma$, where $\sigma =$ 15, 4.4\,mJy beam$^{-1}$ for SO and DCO+ respectively. The magenta solid line is the {\chni infall} profile with conserved angular momentum, and the white solid line is the Keplerian rotation profile with central stellar mass of 0.2 $M_\odot$ {\chni and an inclination angle of $55^\circ$}.
}
\label{fig15}
\end{figure}

\begin{figure}[h]
\centering
\includegraphics[width=\hsize]{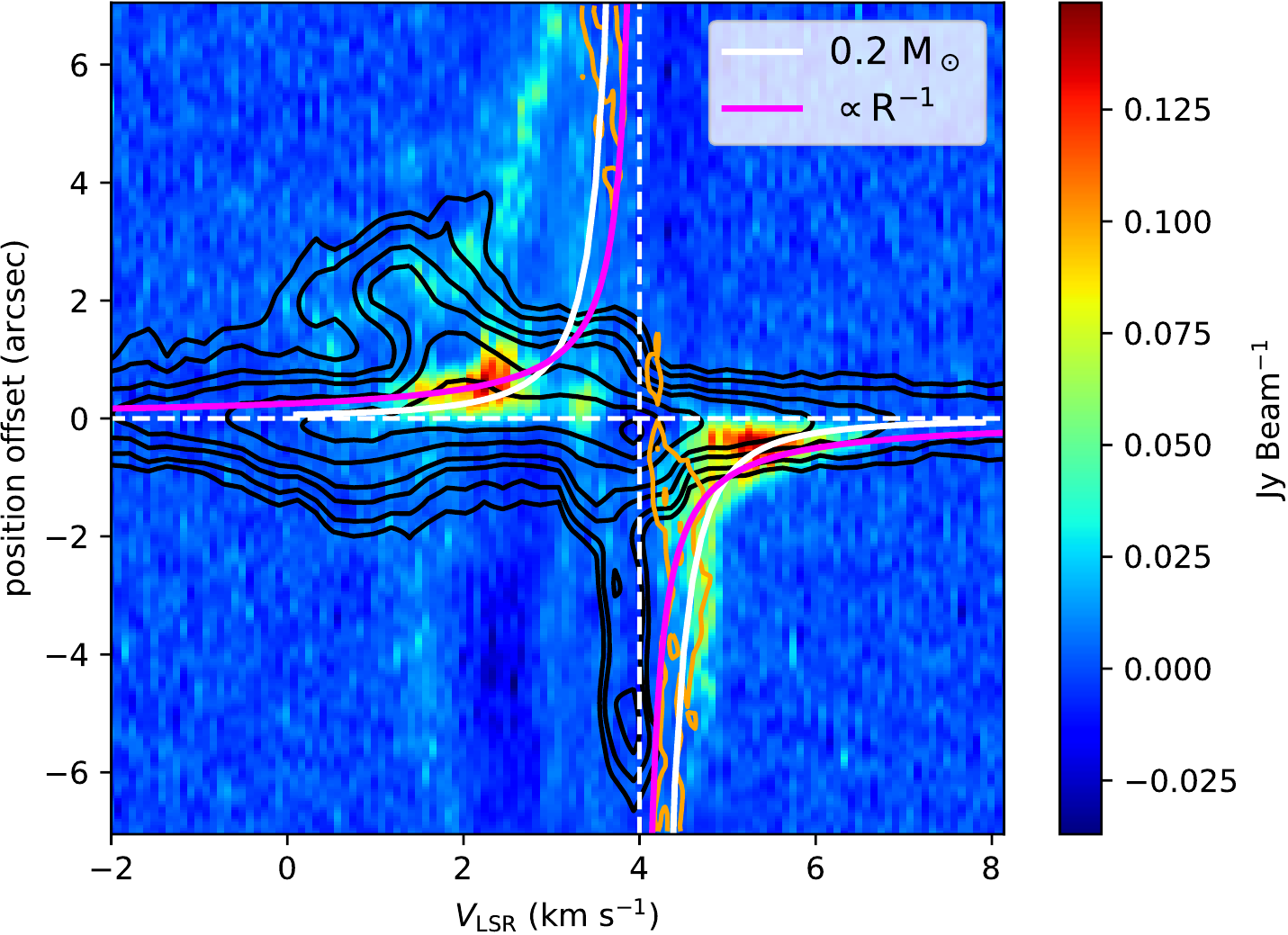}
\caption{ALMA Cycle 4 SO ($\nu=0, J = 8_{8}-7_{7}$, black contours), ALMA Cycle 2 C$^{18}$O {\chni with natural weighting} (J = 2-1, color), and DCO+ (J = 3-2, orange) position velocity (PV) diagram on {\bf VLA1623B}. The contours are in steps of 5\,$\sigma$, 10\,$\sigma$, 20\,$\sigma$, 30\,$\sigma$, 50\,$\sigma$, 70\,$\sigma$, 90\,$\sigma$, where $\sigma =$ 15, 8\,mJy beam$^{-1}$ for SO and DCO+ respectively. The magenta solid line is the {\chni infall} profile with conserved angular momentum, and white solid line is the Keplerian rotation profile with central stellar mass of 0.2 $M_\odot$ {\chni with an inclination angle of $55^\circ$}. Note that the central velocity of SO is shifted to 3\,kms$^{-1}$.}

\label{fig16}
\end{figure}

After identifying the corresponding SO streams around VLA1623A circumbinary disk by comparing with the C$^{18}$O data, PV diagrams are used to further study their interactions. We plot the PV diagrams of SO ($J = 8_{8}-7_{7}$), C$^{18}$O (J = 2-1), and DCO$^{+}$ (J = 3-2) {\chs across} VLA1623A circumbinary disk in \autoref{fig15}. The PV cut is {\chni aligned} along the red line shown in \autoref{fig2}. The black SO contours marked out the extended high velocity SO emission on the blue-shifted side. It spreads out from -4\as0 $\sim$ 2\as0 with the center of the circumbinary disk positioned at 0\as0. The spatially extended high velocity SO on the blue-shifted side suggests there are {\cht mild} accretion shocks, which are likely produced by the interaction between the accretion flow {\cht I, III} and the circumbinary disk. {\chni Since the northern SO stream is perpendicular to the outflow direction, the} contribution from interaction with an outflow {\chni is} ruled out. In contrast, the SO on the red-shifted part is very spatially compact and locates only in the center of the circumbinary disk. The compact structure of the red-shifted SO suggests that there is no violent accretion shocks between the red-shifted accretion flow {\che VI} and the circumbinary disk around VLA1623A. 


\begin{figure*}[tbh]
\centering
\makebox[\textwidth]{\includegraphics[width=\textwidth]{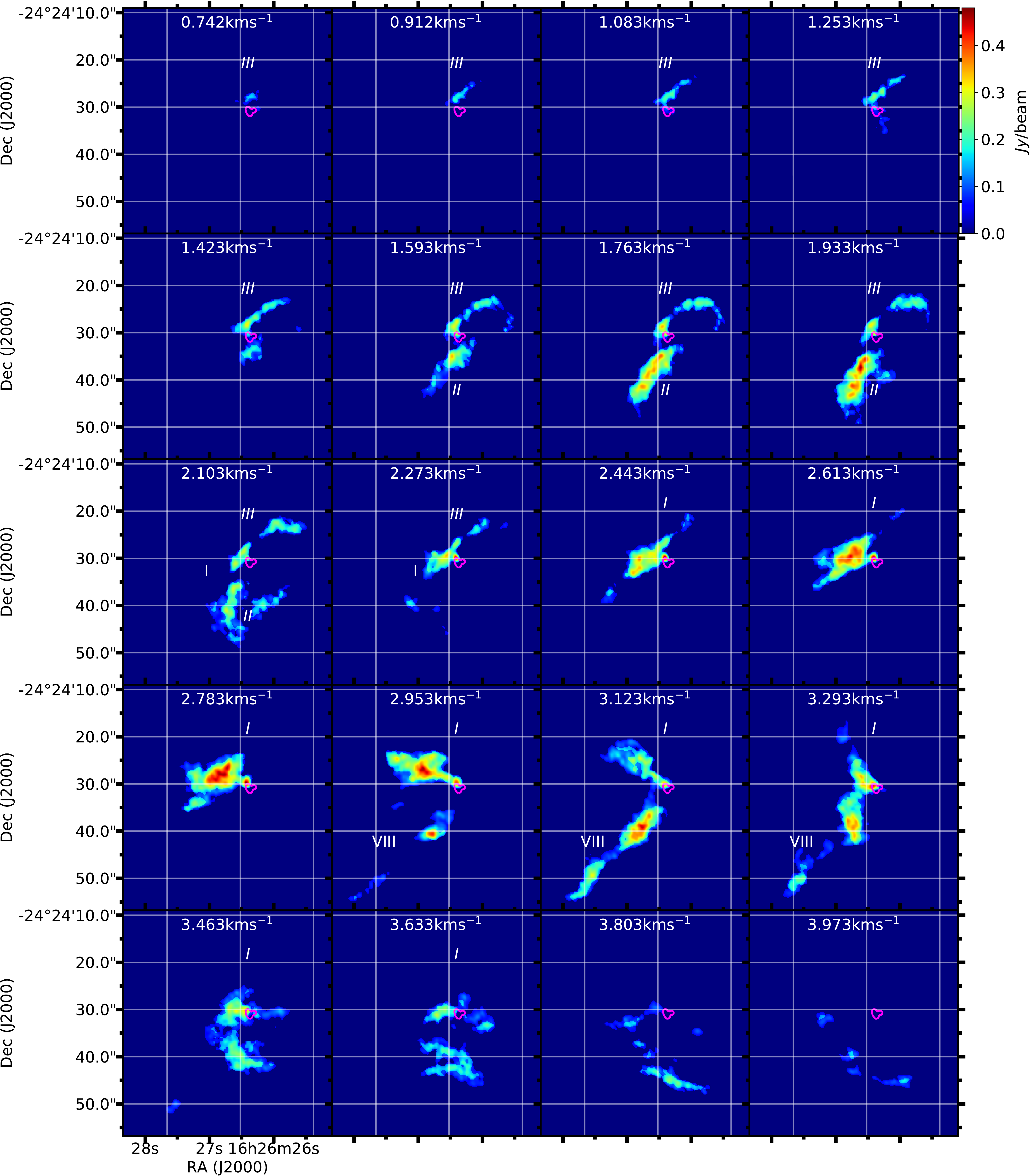}}
\caption{{\cht VLA1623A and blue-shifted {\chel structure} I, II, III and VIII channel maps. The color represents the C$^{18}$O J = 2-1 (Briggs -1.5) emission identified by the dendrogram algorithm. The magenta contours are 0.88 mm continuum data. The contours are in step of 5\,$\sigma$, where $\sigma = 5\times 10^{-4}$ Jy beam$^{-1}$.}
}
\label{fig17}
\end{figure*}

\subsection{{\cht Outflow signatures from VLA1623A and VLA1623B}}

{\cht In \autoref{fig17} we plot the channel maps of the four large scale blue-shifted structures from the dendrogram analysis. In previous sections, based on the CO outflows and SO shocks in \autoref{fig13}{\chel ,} we have established that structure II is an outflow cavity wall. At the exact same position as structure II in \autoref{fig17}, we discovered a similar elongated structure (VIII) at lower velocity ($2.953\sim 3.293\,km\,s^{-1}$). Structure (VIII) has the same shape and position as structure II, and it is also on top of both the CO outflow and SO shocked southern stream suggesting it is also an outflow cavity wall. The two outflows cavity walls traveling at different velocities is a {\chel strong evidence indicating} that there are two outflows in the plane of sky, coming from VLA1623A and  {\chel VLA1623B, respectively.} Cycle 0 CO results \citep{2015A&A...581A..91S}, which is almost completely filtered out by ALMA, suggests that VLA1623B is driving a much more compact {\chel outflow,} and the authors associate the slower large-scale outflows with VLA1623A. 

In contrast, our high resolution Cycle 6 CO data and the discovery of outflow cavity walls (Structure II and VIII) suggests otherwise. There are two outflows overlaying on top of each other in the plane of sky. The large scale outflows come from both VLA1623A and VLA1623B as shown by the two outflow cavity walls at two different velocities. From our Cycle 6 CO data we found the outflow from VLA1623B is more red-shifted {\chel compared} to VLA1623A, therefore we associate the outflow cavity II with outflows from VLA1623A and outflow cavity VIII with outflows from VLA1623B. Multi-tracer analysis to distinguish between the two outflows would be {\chel presented} in a future paper.}

\subsection{Existence of Wall-like structure south of VLA1623B }
\label{sec:4.6}

The PV diagrams of SO ($J = 8_{8}-7_{7}$), C$^{18}$O (J = 2-1), and DCO$^{+}$ (J = 3-2) on VLA1623B, {\che centered at $\alpha$(J2000) = 16\textsuperscript{h}26\textsuperscript{m}26\fs305, $\delta$(J2000) = --24\degree24\arcmin30\as705 with position angle of 222.8$^{\circ}$} are plotted in \autoref{fig16}. In the blue-shifted region at the position offset around 3\as0, we observed the similar extended high velocity blue-shifted SO feature {\chni (compare to \autoref{fig15})}, which corresponds to the shock fronts of the accretion flow {\cht I and III}. The SO emission has a very wide line width ($>$10 kms$^{-1}$) at the position offset between $\pm $1\as0 (across VLA1623B). The huge velocity dispersion on VLA1623B indicates there is a huge change in velocity on VLA1623B and huge shock fronts {\ch are formed}. Furthermore, at the south of VLA1623B the materials only have velocity around 4.0 kms$^{-1}$ suggesting the SO is at rest. The huge change in SO velocity and materials (C$^{18}$O, SO, DCO$^{+}$) on the south of the VLA1623B {\chs are} at rest both suggests a wall-like structure is located south of VLA1623B. 

In {\chs the} previous section, the {\chs red-shifted} accretion flow {\che VI} is identified to be connecting to VLA1623B {\chs (See \autoref{fig6},\, \autoref{fig8})}. When the materials from the {\chs blue-shifted} accretion flow {\chs III} accrete onto VLA1623B, {\ch they are} quickly stopped by the red-shifted accretion flow {\che VI} (shown in \autoref{fig8}). The collision between {\chs blue-shifted} accretion flow {\chs III} and {\chs red-shifted} accretion flow {\che VI} on VLA1623B slows down the materials and forms an extended wall-like structure south of VLA1623B. This explains why no violent accretion shocks from the red-shifted accretion flow {\che VI} is observed around VLA1623A circumbinary disk. The {\chf red-shifted} accretion flow {\che VI} is already significantly slowed down around VLA1623B.

{\chf To constrain the size of the wall-like structure south of VLA1623B, we analyze the SO PV diagram in \autoref{fig16}. In \autoref{fig16}, around the systematic velocity 4 kms$^{-1}$ there exists a long extended SO and DCO$^{+}$ on the south (negative offset) side of VLA1623B. DCO{\cht $^+$} would have abundance enhancement when the temperature is below CO freeze-out temperature  \citep{2013A&A...557A.132M}. {\ch However, DCO{\cht $^+$} emission at rest around the disk is contaminated} from the envelope {\ch making it not ideal to trace the wall-like structure south of the circumbinary disk.} {\ch On the other hand, SO which has high sublimation temperature of 50K traces the shocks region near the centrifugal barrier \citep{2014Natur.507...78S}.} The SO in \autoref{fig16} extend{\ch s} to around 6\as5 ($\sim$ 780 AU). Therefore, the wall on the south side of VLA1623B {\chs has a plane of sky width} at least 780\,AU.}

\section{Discussions}
\label{sec:discussion}

\subsection{{\cht Explanations of the super-Keplerian rotation in the inner region of the disk}}

\begin{figure*}[tbh]
\centering
\makebox[\textwidth]{\includegraphics[width=\textwidth]{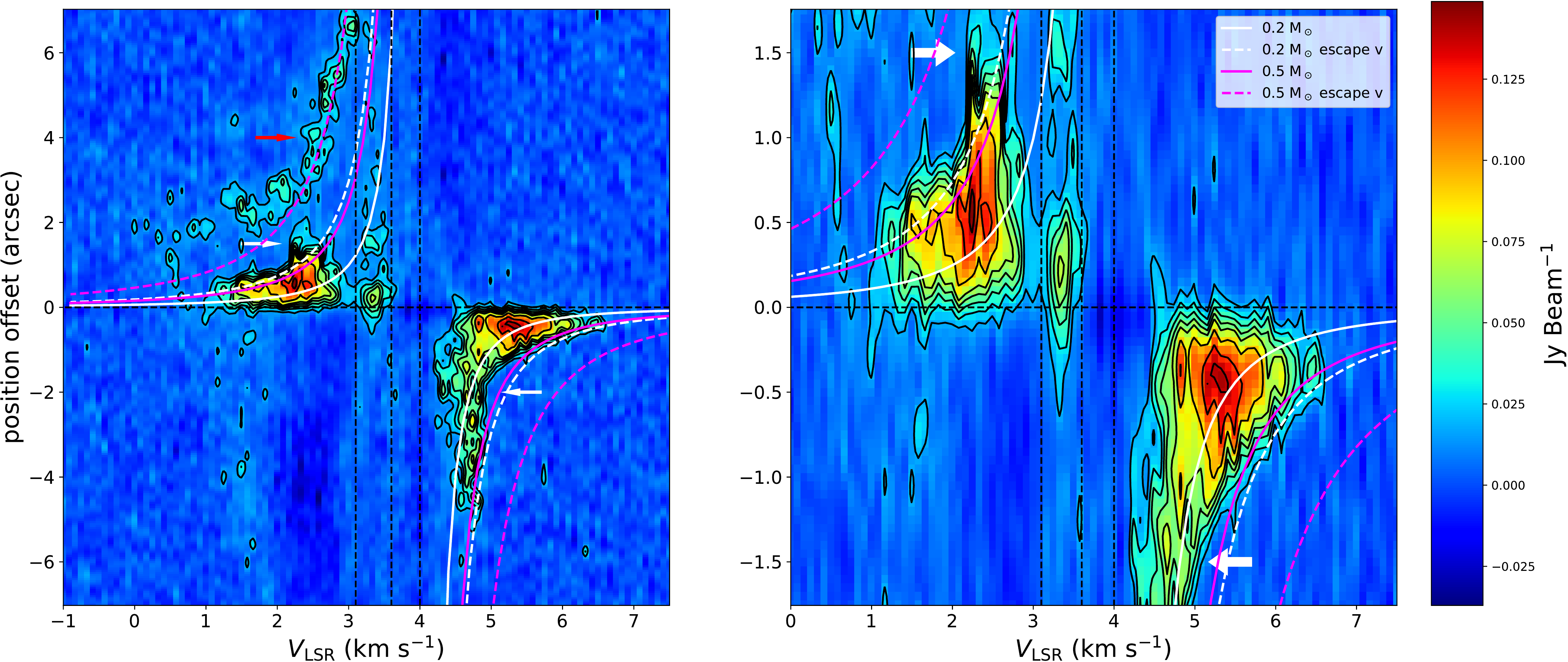}}
\caption{{\chni (a). Position-Velocity (PV) cut of C$^{18}$O J = 2-1 ALMA Cycle 2 VLA1623A Keplerian disk (natural weighting), centered at VLA1623A with position angle 209.82$^{\circ}$. The white and magenta solid lines represent the Keplerian rotation profile with a central mass of 0.2 $M_{\odot}$ and 0.5 $M_{\odot}$ respectively. {\chni The inclination angle is set to be $55^\circ$.} The dotted lines marked their corresponding escape velocity. {\che The white arrow marks the position of the Keplerian disk that is asymmetric in motion. The red arrow marks the blue-shifted accretion flow I. The vertical black lines mark the velocities of the major {\cht large scale emission}.} (b). {\che The zoomed} in image of (a).
}}
\label{fig18}
\end{figure*}

As discussed in {\cht \S~3.2}, we identified a blue-shifted {\che (super-Keplerian)} rotation region between 1.5 \,kms$^{-1}$ and 2.0 \,kms$^{-1}$ within 1\as0 in position offset. A clear gap {\che between VLA1623A circumbinary disk and {\chs the blue-shifted accretion flow I} can be found between 2.3\,km\,s$^{-1}$ to 3.2\,km\,s$^{-1}$ in \autoref{fig7} as well as the PV diagram in \autoref{fig4} a. marked by a white arrow.} This gap sets {\che a clear} boundary between {\chs blue-shifted} accretion flow {\chs I} and {\chs the} disk. {\che This further rules out the possibility that the blue-shifted super-Keplerian rotation region is part of any large scale structures or accretion flows in the line of sight as the disk and accretion flow I are clearly separate in position and velocity space by a gap is shown in \autoref{fig4} a.} {\che The fact that magenta {\chni infall} velocity profile passes through the inner region of the disk, but deviates significantly on the outer edge of the circumbinary disk on the red-shift side suggests this high-velocity super-Keplerian region inside the disk has different angular momentum from the large scale accretion flows.} Thus, an important question remained to be answered is whether or not the high-velocity blue-shifted component (super-Keplerian rotation region) is part of the disk structure? 

One possible explanation for this super-Keplerian rotation region is disk flaring. For a flared disk, due to the z-direction projection effect it is possible the inner region of the disk is projected to {\chni positions} further away from the disk center. To take into account of disk flaring and projection effects, we developed a more sophisticated 3D {\che flared} disk model to model the observation data. By comparing the model with the ALMA data, we {\chel tested this interpretation}.

{\cht As shown in \autoref{fig21},} all the flared disk models (thick black contours) can't explain the blue-shifted super-Keplerian rotation region in 0\as5 position offset at velocity range 1.5 $\sim$ 2.0 kms$^{-1}$. The mismatch between the flared disk model and the data shows that the super-Keplerian rotation region is not {\che due to projection effects} of a flared disk. If the super-Keplerian region is due to projection effects, one would expect the disk to be very flared, so the higher velocity materials in the inner region can be projected at larger position offsets. In \autoref{fig9} {\cht at 1.5\,km\,s$^{-1}$} the white Keplerian rotation line {\cht has a}  position offset of 0\as2 ($\sim$ 24\,AU) {\cht lower than} the {\cht C$^{18}$O data which} locates at 0\as5 $\sim$ 0\as7 (60 $\sim$ 80\,AU){\cht . Considering} an inclination of 55$^{\circ}$, if the super-Keplerian rotation region is due to project effects, then one would expect the majority of the C$^{18}$O is distributed around 40$\sim$70\,AU above the disk plane. To achieve this, the scale height of the disk, location where density drops to a fraction of 1/e from the mid-plane, must be much greater than 40\,AU. {\cht Our} flared disk models constrained the disk scale height to be within 15\,AU, {\chel and} thus shows this super-Keplerian rotation region is not due to projection effects from a flared disk.}  




The flat and flared Keplerian disk modeling cannot explain the high-velocity super-Keplerian rotation region in the inner part of the disk. Previous dendrogram {\chni analyses found a} gap between accretion flows and circumbinary disk suggesting that this super-Keplerian rotation region is not coming from large scale accretion flows in the line of sight. It is within the 180 \,AU of the circumbinary disk. There are two possible explanations to this super-Keplerian rotation region: (i) {\cht It is due to collision with infalling materials from the envelope. The  materials from  above the circumbinary disk plane fall onto the inner region of the circumbinary disk. The collision resulted in the net gain of angular momentum in the inner region of the disk}. (ii) {\cht Previous Cycle 0 data underestimate the mass, the combined mass of the binary should be {\chel 0.3}$\sim$0.5\,$M_\odot$ instead of 0.2\,$M_\odot$.}  

{\cht The collision between the circumbinary disk and the infalling materials from above the disk plane can provide enough acceleration to explain the super-Keplerian rotation region}. Previously, we have used the dendrogram to identify large scale accretion flows. We found a 120 AU wide gap between the accretion flow {\chs I} and the circumbinary disk. In order to create high-velocity C$^{18}$O component only in the inner region of the circumbinary disk, the {\chni infalling} materials can only collide with the circumbinary disk from above the disk plane. {\cht Moreover, with outflows perpendicular to the disk the allowed infalling angles lie only between the outflow and the disk. Note that one important feature of the super-Keplerian region is that it is symmetric in both blue and red-shifted region. Consider the materials moving in a 60\,AU circular orbit with circular velocity on the order of $\sim 2$\,kms$^{-1}$, the time scale to complete 1 orbit is on the order of $\sim$200 years. This time scale is significantly smaller than the disk evolution time scales. We expect any asymmetry features caused by the infalling materials onto the disk to be smoothed out, creating a symmetric feature on both blue and red-shifted side of the PV diagram. The short orbital period compared to {\chel the outer} disk evolution timescales ($\sim 10^4$ years) also implies any infalling signatures are only transient structures making them unlikely to be observed if the infalling materials are not continuously supplied by the envelope.} 

{\cht Another possible explanation of the inner super-Keplerian region is a higher combined binary mass. This would mean the previous mass estimate (0.2 $M_{\odot}$) from Cycle 0  \citep{2013A&A...560A.103M} underestimates the mass. To check this possibility, we plot the escape velocities as dotted lines for both combined binary mass of 0.2 $M_{\odot}$ and 0.5 $M_{\odot}$ in \autoref{fig18}.}


The upper left stream (Blue-shifted accretion flow I marked by the red arrow) appears to match the escape velocity for 0.5 $M_{\odot}$. {\chni Infalling} streams must have velocity less than the escape velocity. Thus, the sum of {\cht surrounding envelope} mass and combined binary mass must be greater than 0.5 $M_{\odot}$. Since Class 0 protostar is still deeply embedded in the {\che envelope} \citep{2015MNRAS.446.2776S}, and envelope mass could be comparable or larger than the central star, the motion of accretion flows cannot completely break the degeneracy in combined binary mass. 

\begin{figure*}[tbh]
\centering
\makebox[\textwidth]{\includegraphics[width=\textwidth]{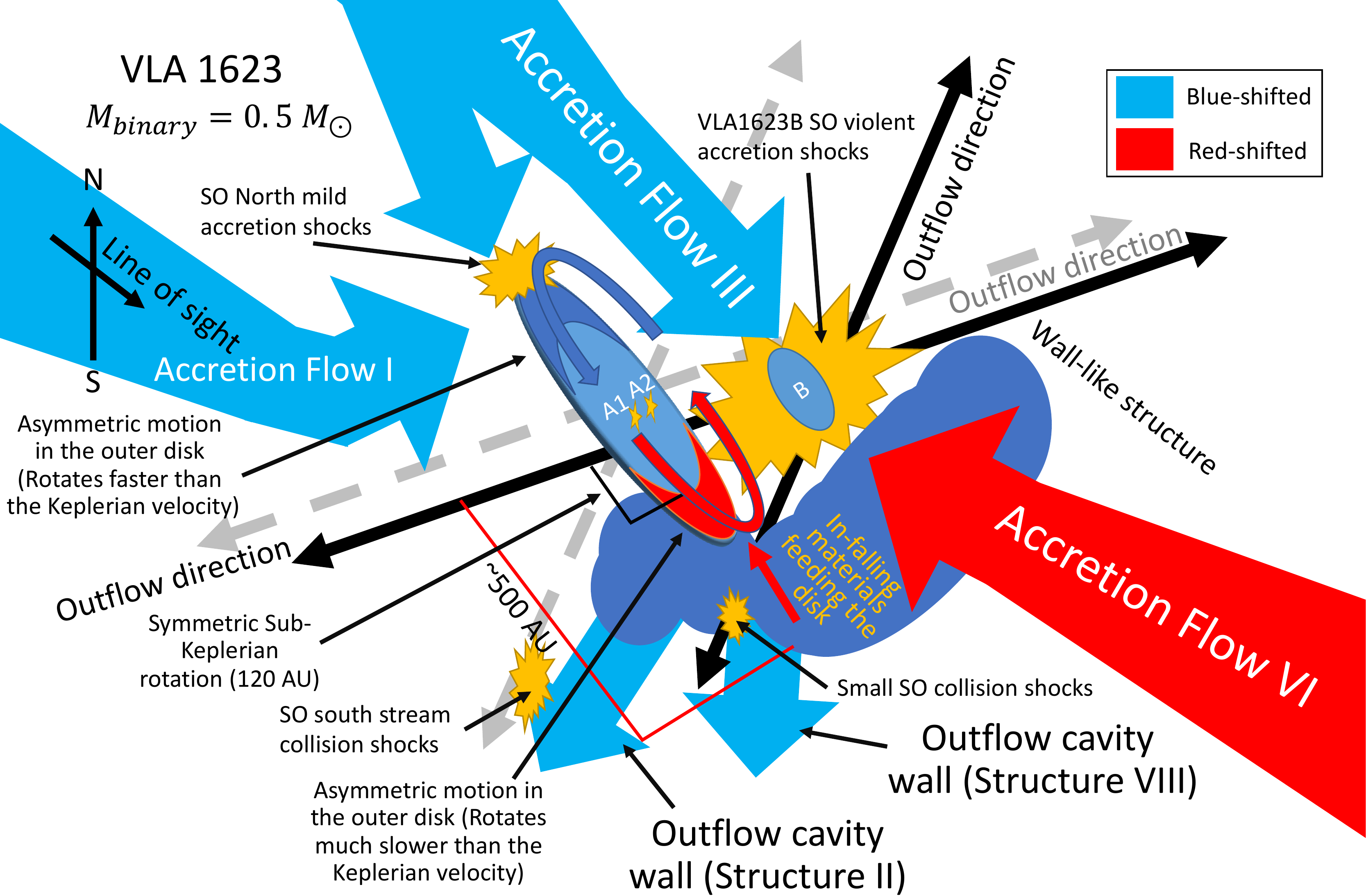}}
\caption{\cht Cartoon diagram of accretion flows and disk interactions studied in this work towards VLA1623A circumbinary disk and VLA1623B for \textbf{combined VLA1623A binary mass 0.5\,M$_\odot$.}
}
\label{fig19}
\end{figure*}

{\chni Unlike the case with the combined binary mass of 0.2 $M_{\odot}$ \citep{2013A&A...560A.103M} (white solid line \autoref{fig18}), the Keplerian rotation line for combined binary mass of 0.5 $M_{\odot}$ (magenta solid line \autoref{fig18}) passes through the high velocity blue-shifted component perfectly. It will however need the circumbinary disk to be asymmetric in the outer region. For the magenta solid lines, the blue-shifted disk between 1\as0 to 2\as0 rotates faster than the Keplerian rotation, while at the same position the red-shifted disk is sub-Keplerian (marked by white arrows). The observation of blue-shifted SO shocks north of the disk might be a possible explanation of this asymmetry in motion. 

{\cht To break the mass degeneracy, we fit the PV diagram across VLA1623A with {\chel a} careful treatment. First, we masked out all pixels with negative flux. Then for each velocity channel we search for the peaks above 3 sigma at different position. This will allow us to identify different structures at each velocity channels. Then we remove the infalling stream (red arrow in \autoref{fig18}) and uses the data that are at least 0.5\,kms$^{-1}$ away from the systematic velocity (4.0\,kms$^{-1}$) to prevent contamination. Intensity weighted position average is then calculated to determine the representative disk position for each velocity channel. The non-linear least squares fitting result yields a central binary mass of $0.3$\,M$_\odot$.} {\cht If one assumes that all gas should be sub-Keplerian in the disk and the Keplerian rotation line should ``match only the edge of the PV diagram", then a mass of $0.5$\,M$_\odot$ would be obtained. }

{\cht From the flat and flared disk modeling to the infalling streams from accretion flow I, we have concluded that the most plausible explanation of super-Keplerian motion is the underestimation of combined binary mass in VLA1623A. The combined binary mass from VLA1623A should be 0.3$\sim$\,0.5\,$M_\odot$.}



\subsection{{\chni Summarized Picture}}

The results of flat and flared disk modeling show {\che that} VLA1623A circumbinary disk is a large flat Keplerian disk {\che with} a size of 180 AU {\cht and a combined binary mass of 0.3$\sim$\,0.5\,$M_\odot$.} At the edge of the circumbinary disk, it is {\cht estimated} to have a thickness {\cht around 30\,AU based on the of CMU modeling}, which shows the thickness of the incoming accretion flows at the centrifugal radius is around 30 AU (Cheong et al. {\chni 2019; submitted}). 

In the previous sections, we used both SO $J = 8_{8}-7_{7}$ and C$^{18}$O J = 2-1 data to study how the accretion flows interact with the circumbinary disk around VLA1623A and VLA1623B. A cartoon diagram of their interactions is summarized in  {\cht \autoref{fig19} for combined binary mass  0.5\,M$_\odot$}. In short, there are around {\chni 3} main accretion flows found in this study: blue-shifted accretion flow I, III, and red-shifted accretion flow VI {\che as} summarized in \autoref{table:3}. 


\begin{table}
\setlength{\tabcolsep}{4.6pt} 
\caption{Accretion flows around VLA1623A}             
\label{table:3}      
\centering                          
\begin{tabular}{c c c c}        
\hline\hline                 
{\chni Structure}  & Velocity & SO data & C$^{18}$O data  \\    
 	   & (kms$^{-1}$) &     & 				  \\
\hline                        
   {\chni Accretion}  & 2.02 $\sim$ 3.60 & \autoref{fig11} {\chni East}   &  \autoref{fig7}\\  
      {\chni flows I}  	   &  & SO stream     &  \\
\hline                                   

    {\chni Outflow} &  & \autoref{fig11} & \autoref{fig12} b \\
    {\chni cavity wall}  & 1.27 $\sim$ 2.21   &South SO    & South C$^{18}$O\\
    {\chni II}   &				& stream 	 &  stream\\
\hline                                   
   {\chni Accretion} &  & \autoref{fig11}    & \autoref{fig12} b  \\
   {\chni flows III} & 1.27 $\sim$ 2.21  & North SO  & North C$^{18}$O \\
            & & stream & stream \\

\hline                                   
   {\chni Accretion}    	&  				&  \autoref{fig11}& \autoref{fig8}, \ref{fig12}  a\\
   {\chni flows VI}          & 3.45 $\sim$ 4.79 &  {\chni West} SO   &  {\chni West}\\
                &  				&  stream       & C$^{18}$O clump\\

 \hline                                   
   {\cht Outflow}    	&  				& \autoref{fig14} &  \\
   {\cht cavity wall}   & {\cht 2.95 $\sim$ 3.29} & {\cht channel} & \autoref{fig17} \\
   {\cht VIII}          & & 	{\cht  3.084\,km\,s$^{-1}$}   &  \\
                
\hline\hline                                 
\end{tabular}
\end{table}

From the extended emission in the SO PV diagram (\autoref{fig15}), we identified an accretion shock north of the circumbinary disk around VLA1623A. This SO accretion shocks are produced by the {\chs blue-shifted} accretion flows {\chs I} and {\chs III} colliding with the edge of the circumbinary disk. The {\chs blue-shifted} accretion flow {\chs III} also collide with the red-shifted accretion flows {\che VI} on VLA1623B (\autoref{fig16}). The collision creates extremely wide SO line width ($>$ 10 kms$^{-1}$) corresponding to the violent shocks on VLA1623B. The collision significantly slows down and stop the materials from the red-shifted accretion flows {\che VI} at a position south of the VLA1623B forming a wall-like structure as shown in \autoref{fig16}. 

The materials from red-shifted accretion flow {\che VI} continue to pile up, spread to the south of VLA1623A circumbinary disk and infall towards it. The {\cht infall of} rotating materials are connected to the boundary of the disk and extended up to $\sim$ 500\,AU south of the circumbinary disk. This explains the extended red-shifted C$^{18}$O emission in \autoref{fig3}. {\chni Furthermore, the outflow collides with this {\chni infall} rotating materials and forms a long extended SO South stream in \autoref{fig13} with a peak located around 4\as0 south from the disk.} The overall picture of the interactions between accretion flows and circumbinary disk around VLA1623A and VLA1623B is summarized in {\cht \autoref{fig19}}. 

\subsection{{\chni SO North of the VLA\,1623A circumbinary disk and VLA\,1623B; shocks or infall/rotation?}}

The SO north of the disk and VLA1623B in \autoref{fig15} and \autoref{fig16} {\chel (more positive positional offsets)} appears to have similar linewidth as compared to C$^{18}$O which traces infall and rotation. However, SO north of the VLA1623A circumbinary disk and VLA1623B is actually tracing a mild shock instead of infall and rotation. In the higher 0\as3 resolution SO 3\,$\Sigma$\, $\nu=0$\, 7(8)-6(7) observation (Project code: 2018.1.00388.S, PI: Cheng-Han Hsieh), 2 compact SO peaks are observed north of VLA1623A and VLA1623B (Hsieh et al. 2019 in prep). The SO linewidths of those peaks are 4\,kms$^{-1}$ and 7\,kms$^{-1}$ which are significantly larger than the C$^{18}$O linewidth (Hsieh et al. 2019 in prep) suggesting they are very likely originated from a shock. More in-depth study of shocks around VLA1623 system would be present in the future paper.

\section{Conclusions}
\label{sec:conclusion}

This work presents a detailed analysis of VLA1623A circumbinary disk and VLA1623B. 
The results can be summarized as the following:

\begin{enumerate}

\item The super-Keplerian rotation region inside the VLA1623A circumbinary disk cannot be fitted properly with either Flat or Flared Keplerian disk models with binary mass 0.2\,M$_\odot$. {\cht This is due to the results from Cycle 0 data significantly underestimate the binary mass. Based on accretion streams (\autoref{fig18}) and disk modeling, we suggest the combined binary mass for VLA1623A should be $0.3\sim$0.5\,M$_\odot$}. 



\item From the SO PV diagrams, we detect the existence of a wall-like structure south of VLA1623B. The wall has a plane of sky width of around 780\,AU on the VLA1623B side. Furthermore, {\chni plausible pictures} of how accretion flows interact with VLA1623A circumbinary disk and VLA1623B {\chni are} constructed and shown in  {\chni \autoref{fig19}}. 


\item {\cht From the dendrogram analysis, we discovered two outflow cavity walls (structure II and VIII) at the same position moving at different velocities. This is strong evidence suggesting that there are two large-scale CO outflows in the plane of sky on top of each other originated from VLA1623A and VLA1623B.} 




\end{enumerate}

\begin{acknowledgements}
      CHH and SPL acknowledge support from the Ministry of Science and Technology of Taiwan with Grant MOST 106-2119-M-007-021-MY3. ZYL is supported in part by NASA grant 80NSSC18K1095 and NNX14B38G and NSF grant AST-1815784 {\chni, 1910106, 1716259}. The authors further thank H\'{e}ctor  {\cht Arce, Paolo Coppi and the referee} for helpful discussions and/or for commenting on drafts of this manuscript. {\che This paper makes use of the following ALMA data: ADS/JAO.ALMA \#2011.0.00902.S, \#2013.1.01004S, \#2015.1.00084S, \#2016.1.01468. {\chni \#2018.1.00388.S.} ALMA is a partnership of ESO (representing its member states), NSF (USA) and NINS (Japan), together with NRC (Canada) and NSC and ASIAA (Taiwan) and KASI (Republic of Korea), in cooperation with the Republic of Chile. The Joint ALMA Observatory is operated by ESO, AUI/NRAO and NAOJ. The National Radio Astronomy Observatory is a facility of the National Science Foundation operated under cooperative agreement by Associated Universities, Inc.} 
\end{acknowledgements}

{\cht \software{RADMC-3D (Dullemond 2012), GLUE (Beaumont et al. 2014), CASA (McMullin et al. 2007), astropy (The Astropy Collaboration 2013, 2018).}}

\appendix
\section{{\chni Estimation of Centrifugal Radius}}
\label{sec:appendix1}
{\chni Consider a simple disk such that the gravitational acceleration balances the centrifugal force where $j$ is the specific angular momentum.
   \begin{align}
      \frac{GM_{*}}{r^2} = {\frac{j^{2}}{r^3}}
      \label{eq:A1}
      \end{align}
      
The centrifugal radius $r_c$ can be expressed in terms of the specific angular momentum ($j$) as following,
   \begin{align}
      r_c = {\frac{j^{2}}{G{\cht M_{*}}}}
      \label{eq:A2}
      \end{align}

If the disk is in hydro-static equilibrium in the vertical direction (z), the total column density ($\sigma$) can be expressed in terms of the volume density at mid-plane ($\rho_c$) as following:      
   \begin{align}
      \sigma = \sqrt[]{\frac{2c_s^{2}\rho_c}{\pi G}}
      \label{eq:A3}
      \end{align}
 where $c_s$ is the sound speed. It is simple to show that
   \begin{align}
      \frac{c_s j}{G{\cht M_{*}}}=\frac{c_s \omega_c r^2}{G\sigma \pi r^2}= \frac{\omega_c}{\sqrt[]{2\pi G\rho_c}}
      \label{eq:A4}
      \end{align}

Numerical simulations by \citet{1997ApJ...478..569M} has shown the factor
\begin{align}
\frac{\omega_c}{\sqrt[]{2\pi G\rho_c}} \approx 0.3. 
\label{eq:A5}
      \end{align}
      
Therefore, the centrifugal radius \footnote{{\chni This expression is modified from the Star Fromation lecture notes by Kohji Tomisaka (2007). See http://th.nao.ac.jp/MEMBER/tomisaka/Lecture\texttt{\_}Notes/StarFormation/3/node85.html}} can be expressed as:
\begin{align}
r_c = {\frac{j^{2}}{G{\cht M_{*}}}}= \frac{GM_{*}}{c_s^2}\Big( \frac{c_s j}{G{\cht M_{*}}} \Big)^{2}\approx0.09\frac{G{\cht M_{*}}}{c_s^2}.
\label{eq:A6}
      \end{align}

Using the sound speed relation,
\begin{align}
c_s^2 = \frac{k_B T}{\mu m_p,}
\label{eq:A7}
      \end{align}
where $\mu \approx 2.3$ and consider a typical disk temperature of 30\,K and a combined binary mass of 0.2\,$M_\odot$, we found that the sound speed $c_s\approx$ 0.33\,km\,s$^{-1}$ and the corresponding centrifugal radius is $\sim$ 148\,AU. For simplicity, the centrifugal radius is close to $\sim$ 120\,AU, so we set the crossover point of constant angular momentum with the keplerian curve to be at 1\as0 in position offset, which corresponds to a specific angular momentum of 120\,AU\,km\,s$^{-1}$.}
\pagebreak
\section{{\chni Flared Disk Modeling Results}}
\label{sec:appendix2}

\autoref{fig21} displays the results of the flared disk modeling without zoomed in on the red-shifted components.  

\begin{figure*}[tbh] 
\centering
\makebox[\textwidth]{\includegraphics[width=\textwidth ]{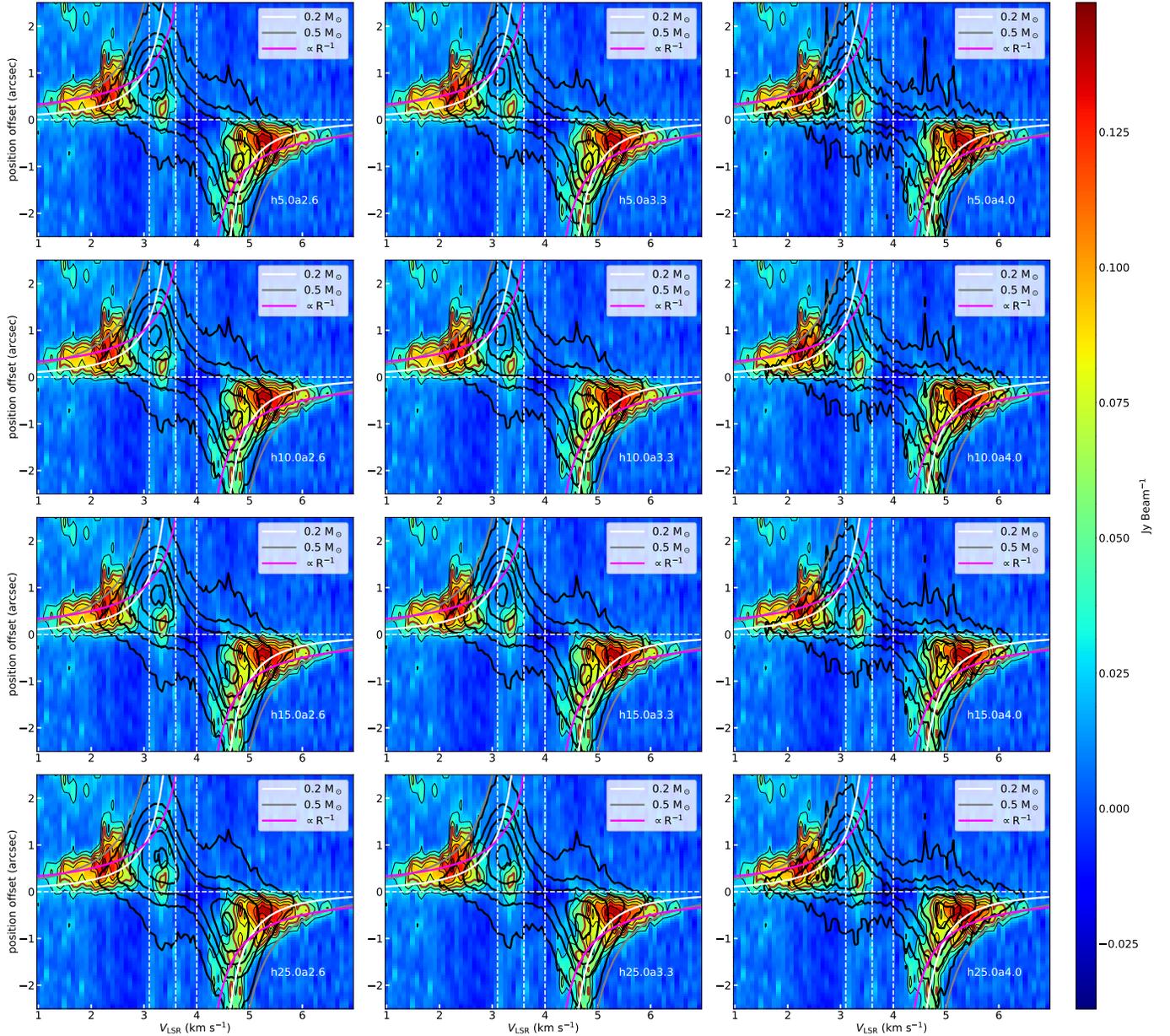}}
\caption{Position-Velocity (PV) diagrams of VLA1623A {\chni keplerian disk overlaid with flared disk model contours} with different vertical scale height $h_{0}$ and density power-law $a$.  The color represents the PV diagram of C$^{18}$O J = 2-1 emission centered at VLA1623A and the thin black contours are plotted for 3\,$\sigma$, 5\,$\sigma$, 7\,$\sigma$, 9\,$\sigma$, 11\,$\sigma$, 13\,$\sigma$, 15\,$\sigma$ with $\sigma=9$\,mJybeam$^{-1}$. The thick black contours are from the Flared Keplerian disk model {\cht with central binary mass 0.2 $M_{\odot}$.} The contours are in steps of 0.2, 0.4, 0.6, 0.8, 0.95 of the maximum flux in {\chf the model}. {\chni The brown contour marks the 7\,$\sigma$ line used for model comparison.} The white {\cht and gray solid lines represent} the Keplerian rotation curve with central star mass {\cht 0.2 and 0.5} $M_{\odot}$ {\chni and an inclination angle of $55^\circ$}. The magenta line represents the {\chni infall} velocity profile with conserved angular momentum. {\chni The vertical white lines mark the velocity of the major {\cht large scale emission}.}
}
\label{fig21}
\end{figure*}
\end{document}